\documentclass{ws-procs961x669}% default, citations in superscript
\newcommand{\pt}{$p_{T}$}
\newcommand{\sq}{$\sqrt{s}$}
\newcommand{\mr}{\mathrm}
\usepackage{lineno}
\begin{document}
%\linenumbers
\title{Soft QCD measurements at LHC}

\author{M. Ta\v{s}evsk\'{y}$^*$}
\begin{center}
  {\em On behalf of the ALICE, ATLAS, CMS, \\
    LHCb, LHCf and TOTEM collaborations}
  \end{center}
\address{Institute of Physics of the Czech Academy of Sciences,\\
Na Slovance 2, Prague, 18221, Czech Republic\\
$^*$E-mail: Marek.Tasevsky@cern.ch}

\begin{abstract}
Results from recent soft QCD measurements by LHC experiments ALICE, ATLAS, CMS,
LHCb, LHCf and TOTEM are reported. The measurements include total, elastic and
inelastic cross sections, inclusive and identified particle spectra, underlying
event and hadronic chains. Results from particle correlations in all three
collision systems, namely $pp$, $pPb$ and $PbPb$, exhibit unexpected similarities.
\end{abstract}


\bodymatter

\section{Introduction}
Soft Quantum Chromodynamics (QCD) physics is a domain of particle physics which is
characterized by a low momentum transfer, typically a low transverse momentum,
\pt. It is usually
used to describe that part of the scattering which dominates at soft scales and
where perturbative QCD cannot be applied. One example of a process which is 
entirely governed by soft QCD physics is the process of hadronization.
Since there is no uniform description of the phenomena that occur at low
\pt, there is a variety of models trying to explain them through comparisons with
extracted data. There is a wealth of LHC measurements that probe the soft QCD
region --- basically all LHC experiments measure soft QCD phenomena. This text
reports on results from experiments ALICE \cite{ALICE}, ATLAS \cite{ATLAS},
CMS \cite{CMS}, LHCb \cite{LHCb}, LHCf \cite{LHCf} and TOTEM \cite{Totem} and
tries to select those which are recent and illustrative at the same time. We will
discuss measurements of inclusive total cross sections,
inclusive and identified particle spectra, underlying event, particle correlations
and it will also be shown that there are surprising similarities between results
from all three collision system: pp, $pPb$ and $PbPb$ collisions. The models which
will be occasionally mentioned are based on multi-parton interactions (MPI),
color reconnections (CR), hadronization and hydrodynamical laws or gluon
saturation in the proton. 
We will also see that there are very interesting links between three big domains
of particle physics, namely particle collisions, heavy-ion collisions and cosmic
rays. They used to be studied separately in the last decades but it turns out
that a wise synergy pays off. 

\section{Total inclusive cross sections}\label{sigmatot}
The total cross section is an important ingredient to estimate the number of
pile-up events at the LHC but also to model interactions in cosmic rays.
The amount of pile-up events, occurring usually at a soft scale, increases with
increasing instantaneous luminosity, and may amount to several tens for values
around the nominal luminosity of $10^{34}$cm$^{-2}$s$^{-1}$. Total,
elastic as well as inelastic cross sections can be measured using special forward
proton detectors which are placed very far from the interaction point and very
close to the beam since their aim is to detect a forward-going proton which
scatters under a very small polar angle. The experiment dedicated for such
measurements is called TOTEM and the special forward proton detector at the ATLAS
side is called ALFA. They are placed more than 200~m from the interaction point
and for the purpose of measuring total and elastic cross sections (which are large
and dominate at low $t$ values, where $t$ is a four-momentum transfer squared)
we use special LHC optics (characterized mainly by the betatron function,
$\beta^*$) and inject only a few bunches with low proton intensity. This leads to
a small number of proton-proton interactions per bunch crossing (often termed
``low pile-up''). The larger $\beta^*$, the lower $t$ values can be reached.
%They are placed more than 200~m from the interaction point
%and they are practically used currently only in runs with special LHC
%optics (characterized mainly by the betatron function, $\beta^*$) which usually
%means colliding only a few bunches of protons, producing low instantaneous
%luminosity and therefore a small number of proton-proton interactions per bunch
%crossing (often termed ``low pile-up'').
\begin{figure}[h]
\includegraphics[height=.25\textheight,width=0.49\textwidth]{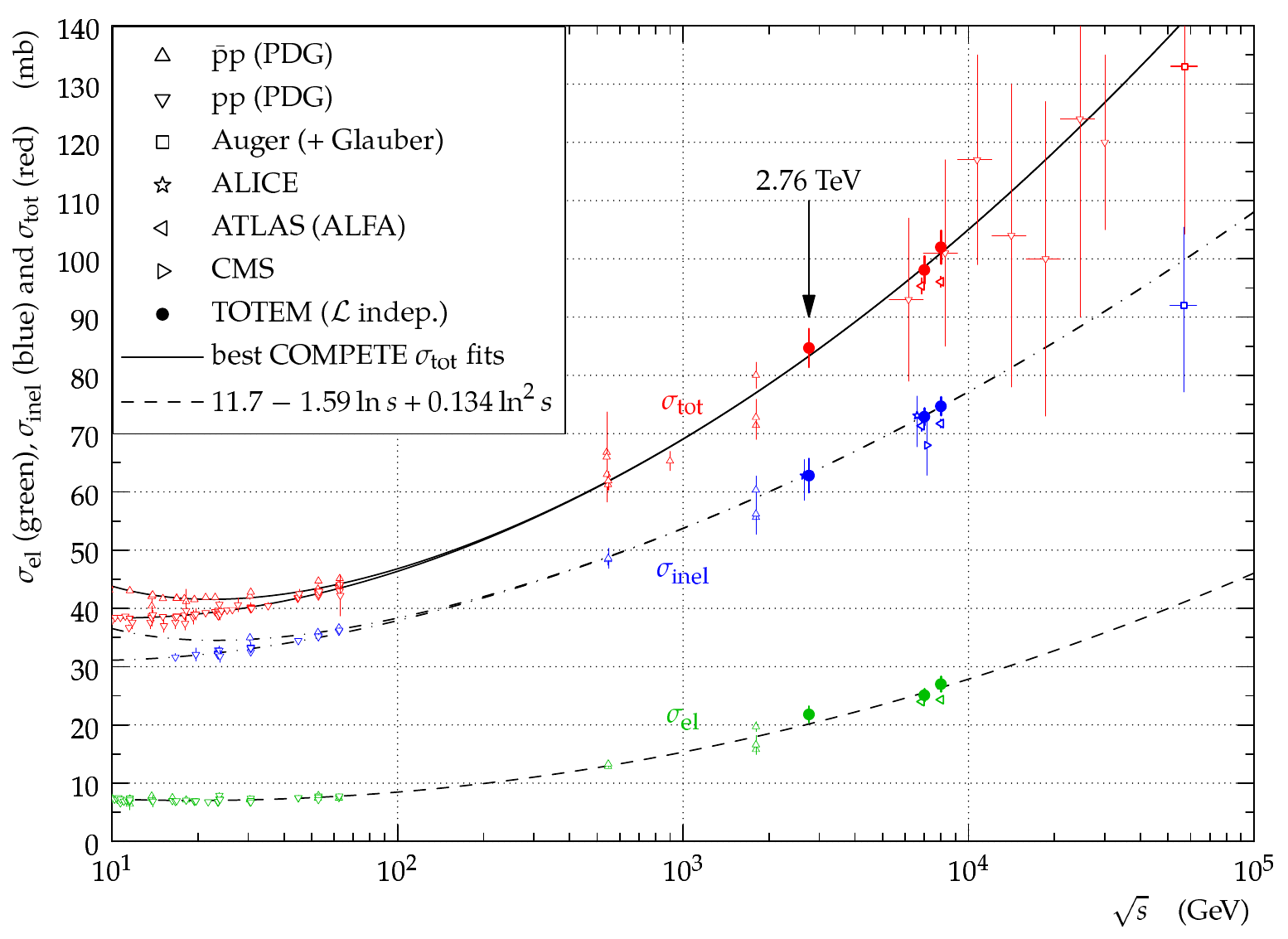}
\includegraphics[height=.25\textheight,width=0.49\textwidth]{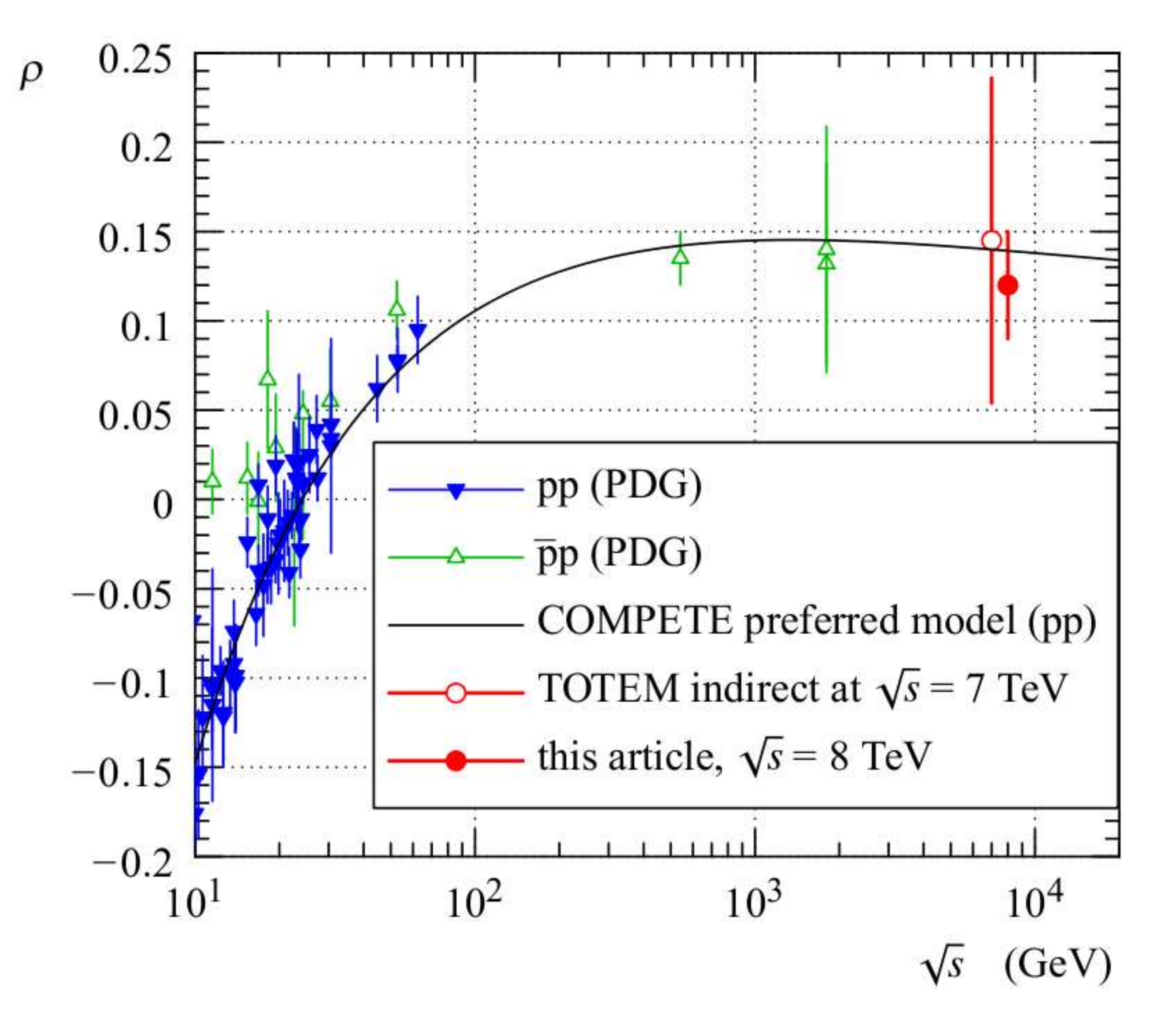}  
  \caption{Left) Compilation of the total, inelastic and elastic cross-section
measurements. The continuous black lines (lower for pp, upper for $\bar{\rm p}$p)
represent the best fits of the total cross-section data by the COMPETE
collaboration \cite{compete}. The dashed line results from a fit of the elastic
scattering data. The dash-dotted lines refer to the inelastic cross section and
are obtained as the difference between the continuous and dashed fits. Right) The
energy dependence of the $\rho$ parameter. The hollow red circle stands for the
earlier indirect determination by TOTEM~\cite{epl101-tot}. The filled red circle
represents the result from TOTEM \cite{Totem-8TeV-sigmatot}. The black curve gives
the preferred pp model by COMPETE~\cite{compete}, obtained without using LHC
data.}
\label{sigmastot}
\end{figure}
Figure~\ref{sigmastot} left shows a compilation of all total, inelastic and
elastic cross section measurements so far together with a preliminary point from
TOTEM \cite{Totem-2.76TeV-sigmatot} for the collision energy, \sq, of 2.76~TeV.
The new ATLAS results for the 8~TeV point \cite{ATLAS-8TeV-sigmatot} are also
included.
%These measurements differ in the LHC optics used for the special runs in which
%the data were collected.
% Mario Deile: vsechny runy krome 2.76 TeV byly beta*=90m
%The larger $\beta^*$, the lower $t$ values can be reached.
The inelastic cross section was recently measured by ATLAS at 13~TeV with the
central detector only \cite{ATLAS-13TeV-inel} and a good consistence was reached
with the results based on the special forward proton detectors as well as with
predictions of PYTHIA~8 \cite{PYTHIA8}, EPOS~LHC \cite{EPOS} and QGSJET-II
\cite{QGSJETII}. 
An interesting feature of the elastic cross section measurement was documented
by TOTEM in the 8~TeV measurement \cite{Totem-nonexp}: in the region where Coulomb
interactions can be safely neglected, the exponential form of the $t$-slope was
excluded at the level of 7.2~$\sigma$. Similar non-exponential $t$-slopes were
also observed at 7 and 13~TeV measurements. The explanation of this
non-exponentiality is still lively discussed among theorists. If $t$ values of the
order of $10^{-4}$ can be reached, one can then study the so called
Coulomb-nuclear interference region which then enables us to measure the $\rho$
parameter which is the ratio of real to imaginary part of the forward amplitude.
Figure~\ref{sigmastot} right shows the \sq\ dependence of the $\rho$
parameter. A direct measurement of the $\rho$ parameter by TOTEM at 8~TeV
($\beta^\star$ = 1~km and $t\sim 10^{-4}$) \cite{Totem-8TeV-sigmatot} agrees within
experimental uncertainties with the COMPETE fit, and another direct measurement,
namely at 13~TeV and $\beta^\star$ = 2.5~km, will appear soon and thus will provide
a valuable level-arm for the energy dependence. 

\section{Inclusive charged particle spectra in pp}
Measurements of inclusive and identified particle spectra belong to basic items
in the physics programs of high-energy experiments. They are usually measured
regularly at each collision energy.
The multiplicity of charged particles is one of the key characteristics of
high-energy hadron collisions and has been the subject of many experimental and
theoretical studies because although quite simple to measure, it is quite
difficult to describe it in the full measured range. Measurements of charged
particle distributions probe the non-perturbative region of QCD where QCD-inspired
models implemented in MC event generators are used to describe the data. 
Measurements are used to constrain free parameters of these models. 
Accurate description of low-energy strong interaction processes is essential for
simulating single pp as well as multiple $pp$ interactions in the same bunch
crossing at higher instantaneous luminosities. Such $pp$ measurements are also used
as input in many models trying to describe heavy-ion results. 
\begin{figure}[h]
\includegraphics[height=.25\textheight,width=0.66\textwidth]{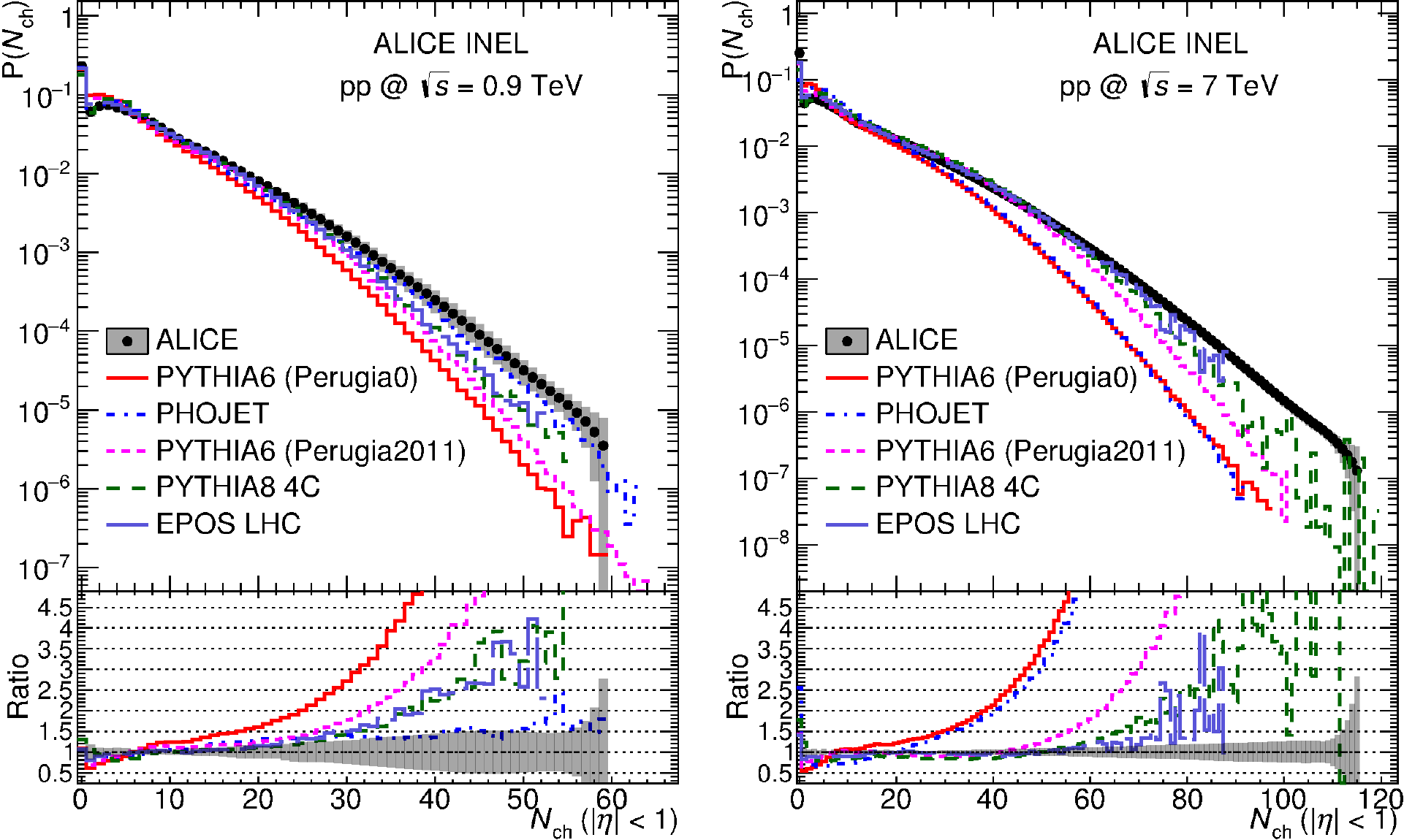}
\includegraphics[height=.26\textheight,width=0.33\textwidth]{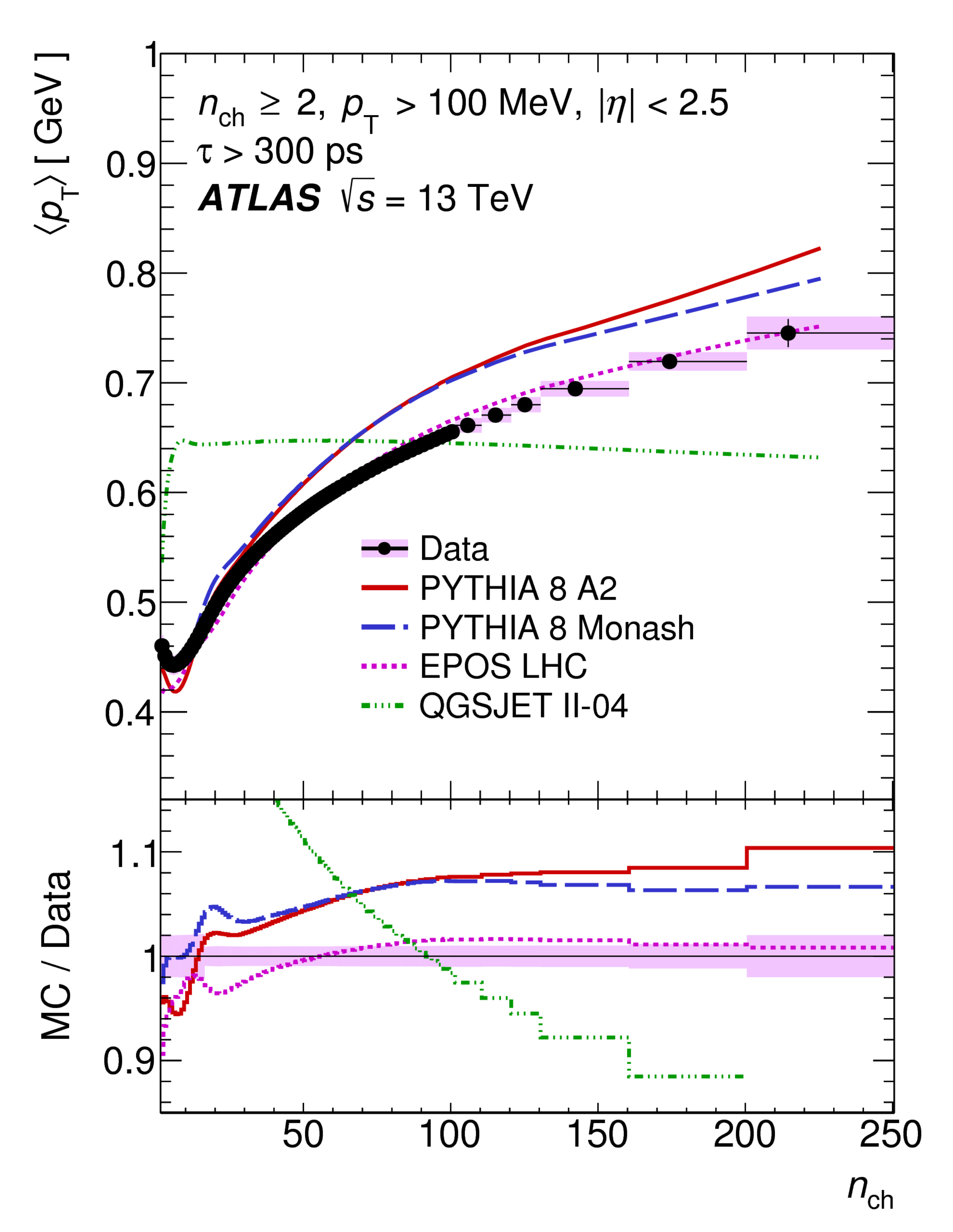}  
\caption{Primary charged-particle multiplicities measured by: ALICE at 0.9~TeV
  \cite{ALICE-incl-ch} (left), 7~TeV \cite{ALICE-incl-ch} (middle) for the INEL
  event class in the pseudorapidity range $|\eta|<$~1.0 and right) by ATLAS at
  13~TeV \cite{ATLAS-13TeV-incl-ch} as a function of the mean \pt\ for events with
  at least two primary charged particles with \pt~$>$~100~MeV and $|\eta|<$~2.5,
  each with a lifetime $\tau >$~300~ps. The black dots represent the data and the
  coloured curves the different MC model predictions. The vertical bars represent
  the statistical uncertainties, while the shaded areas show statistical and
  systematic uncertainties added in quadrature. The lower panel in each figure
  shows the ratio of the data to MC simulation (left and middle) or vice versa
  (right).}
\label{Incl-charged}
\end{figure}
The ALICE analysis \cite{ALICE-incl-ch} presents a comprehensive set of
measurements of pseudorapidity density and multiplicity distributions in $pp$
collisions over the LHC energy range from 0.9 up to 8~TeV, in 5 energy
points. Three event selections are used, namely INEL which means all inelastic
events, then INEL$>$0 which means events with at least 1 charged particle in the
$|\eta|<1$ range, and Non-Single-Diffractive events. Figure~\ref{Incl-charged}
shows the multiplicity distributions at 0.9~TeV (left) and 7~TeV (middle). All
models (EPOS~LHC, PHOJET, PYTHIA~6 and PYTHIA~8) show big troubles in describing
the whole spectrum in data but the best agreement is achieved with EPOS. Similar
observations are made for the remaining \sq\ points.
It is also very useful to measure the energy dependence of the charged particle
multiplicity at mid-rapidity, i.e. for $|\eta| < 0.5$, since it is related to the
average energy density in the interaction of protons and it gives a reference for
heavy-ion collisions. Alternatively, we can calculate normalized $q$-moments
($C_q$) and measure their energy dependence. KNO (Koba, Nielsen and Olesen)
scaling then states that $C_q$ stays energy-independent. Since it was reported in
the past as
dependent on the event selection and $\eta$ range, ALICE came up with a
comprehensive analysis covering three event classes and three $\eta$ ranges. The
conclusion of the measurement is that the KNO scaling violation increases with
increasing \sq\ and at a given \sq, with increasing $\eta$ interval.

A similar study, now at 13~TeV energy, has recently been performed and published
by ATLAS~\cite{ATLAS-13TeV-incl-ch}. Here only charged particles with lifetime
smaller than
30~ps and larger than 300~ps are used since those with lifetime larger than 30~ps
are usually strange baryons which have a low reconstruction efficiency. For
comparison to previous measurements which used also these strange baryons, an
extrapolation is used to include these particles. The data are compared to
PYTHIA~8, EPOS~LHC and QGSJET-II. PYTHIA~8 and EPOS include the effects of color
coherence which is important in dense parton environments and effectively reduces
the number of particles produced in MPI. PYTHIA8 splits the generation into
diffractive and non-diffractive processes, the latter dominated by $t$-channel
gluon exchange, the former is described by Pomeron-based approach. EPOS implements
a parton-based Gribov-Regge theory, effective field theory describing both hard
and soft scattering at the same time. QGSJET-II is based on Reggeon-field theory
framework. EPOS and QGSJET-II do not rely on parton density function (PDF).
Averaged \pt\ as a function of
multiplicity increases as modeled by a colour reconnection mechanism in PYTHIA~8
and by the hydrodynamical evolution model in EPOS. QGSJET-II model which has no
model for colour coherence effects describes the data poorly.
This analysis also reported that multiplicity distribution is not described
perfectly by any of the models, there are large discrepancies especially at large
multiplicities. Having observed similar discrepancies at all measured energies, we
conclude that for every collision energy, model parameters usually need to be re-tuned
in every MC generator.
Unlike the multiplicity distributions, the mean particle multiplicity at
mid-rapidity ($|\eta|<0.2$) measured at several \sq\ points was found to be well
described by PYTHIA~8 Monash and EPOS models for three event selections.

\subsection{Forward energy flow of charged particles in $pp$}
Measurements of the forward energy flow of charged and neutral particles serve to
tune two types of MC models: i) those used at hadron colliders where the
measurements are used to tune MPI and other soft characteristics, ii) the
measurements have also an impact on the total number of muons in the extensive
air showers at the ground whose measurements are still not well-described by
models for cosmic rays.
\begin{figure}[h]
  \begin{center}
\includegraphics[height=.25\textheight,width=0.45\textwidth]{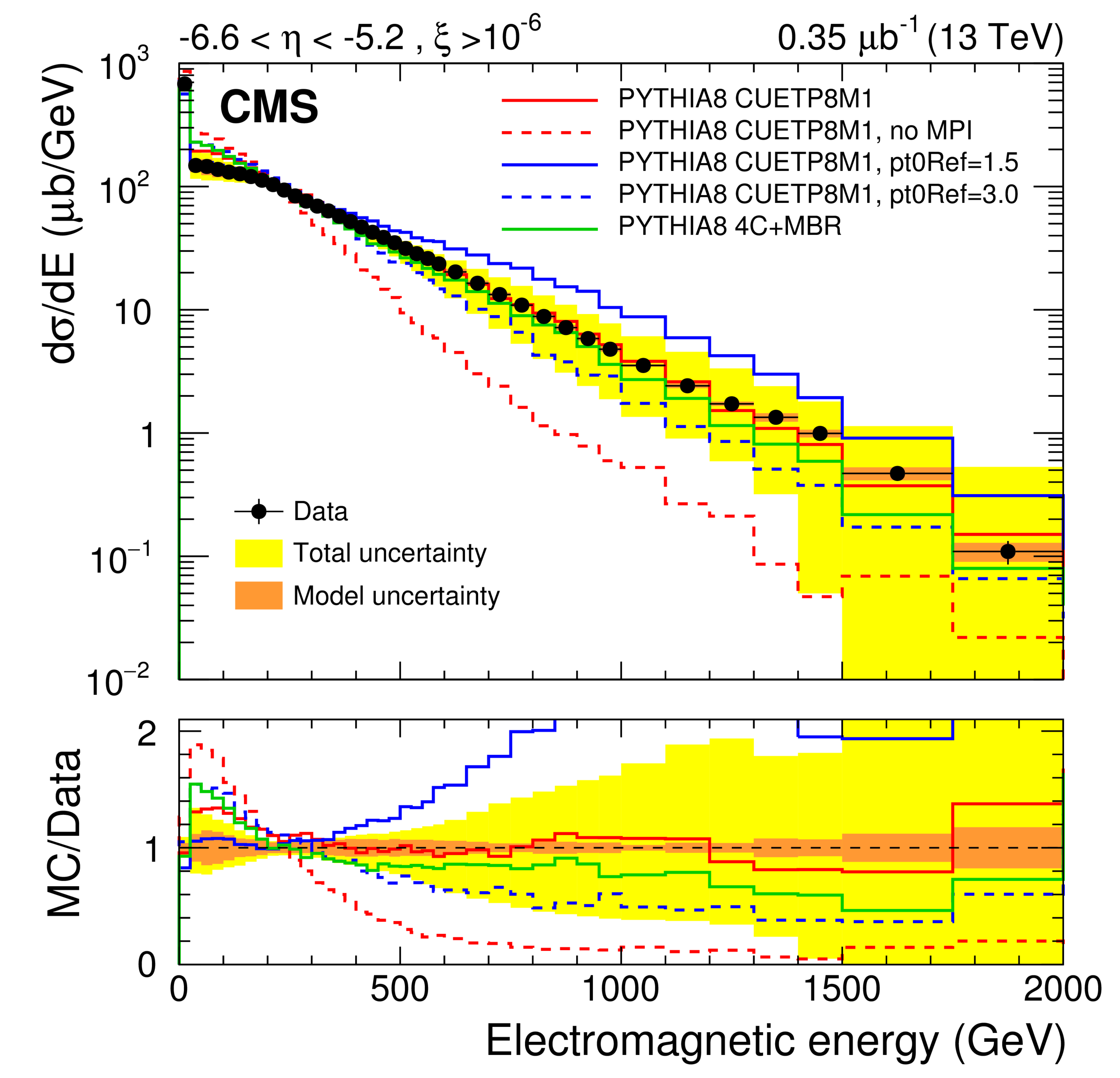}
\includegraphics[height=.25\textheight,width=0.45\textwidth]{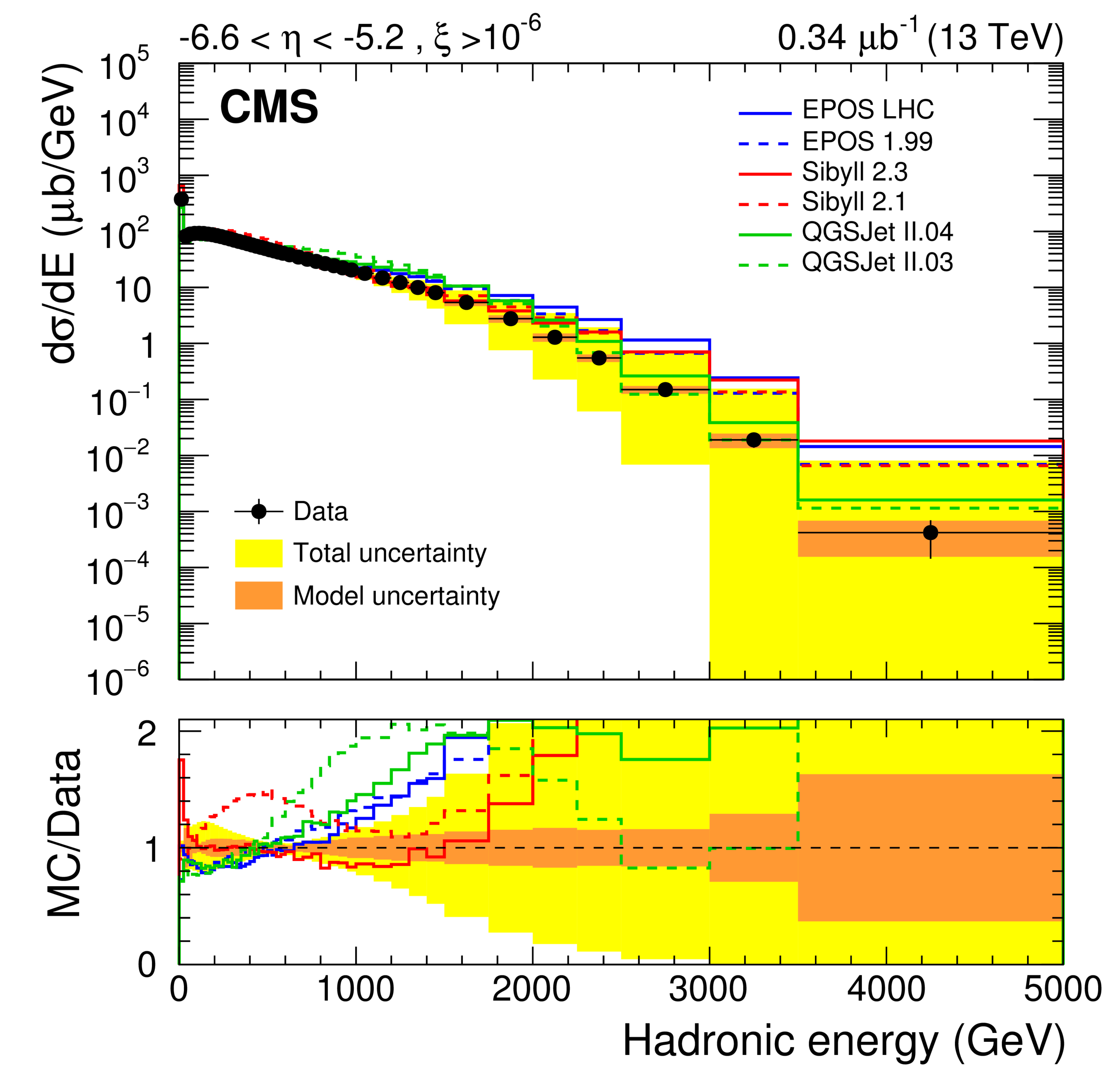}  
\caption{Left) Differential cross section as a function of the electromagnetic
  energy in the region $ -6.6<\eta< -5.2$ for events with $\xi>10^{-6}$ ($xi$ is
  fractional momentum loss of the incident proton). The panel
  shows the data compared to different PYTHIA~8 tunes. Right) Differential cross
  section as a function of the hadronic energy in the region $ -6.6<\eta< -5.2$
  for events with $\xi>10^{-6}$. The panel shows the data compared to MC event
  generators mostly developed for cosmic ray induced air showers. Plots taken from
Ref.~\citenum{CMS-13TeV-CASTOR}.}
\label{CMS-CASTOR}
\end{center}
\end{figure}
The CMS measurement described in \cite{CMS-13TeV-CASTOR} is based on using a
forward calorimeter at negative rapidities only, called CASTOR which
has an electromagnetic and hadronic part, so is able to distinguish electromagnetic
particles (which are mostly electrons and photons from $\pi^0$ decays) and hadrons
(mostly $\pi^{\pm}$). The data were taken at a very low instantaneous luminosity to
suppress the pile-up. In Fig.~\ref{CMS-CASTOR} left, the energy spectra of
electromagnetic particles are compared to predictions of models used to describe
the multihadron production at hadron colliders and from the comparisons to various
MPI tunings, we conclude that the data are very sensitive to the MPI modeling.
The right side of Fig.~\ref{CMS-CASTOR} documents huge deficiencies in
describing the spectrum of hadrons by all models used to model cosmic rays
(EPOS, SYBILL and QGSJET II). 

\subsection{Forward energy flow of neutral particles in $pp$}
The very forward energy flow of neutral particles is measured by a set of
hodoscopes, in a dedicated experiment LHCf located at 140~m downstream of
the ATLAS detector. The aim of such measurements is to improve hadronic
interaction models, especially used in generators for cosmic rays. For example,
modeling of $X_{\mr max}$ (the position of shower maximum) needs the total
cross section for the collision of proton with air, but also the spectra of
identified particles going very forward. And modeling of the hadronic interactions
in those generators is based on correlations between spectra of particles going
central versus those going forward. This is perfectly possible since LHCf uses the
same interaction point as the ATLAS detector which measures very precisely the
central production. A first common analysis of the same event sample analyzed by
ATLAS and LHCf is in preparation.
\begin{figure}[h]
\includegraphics[height=.23\textheight, width=\textwidth]{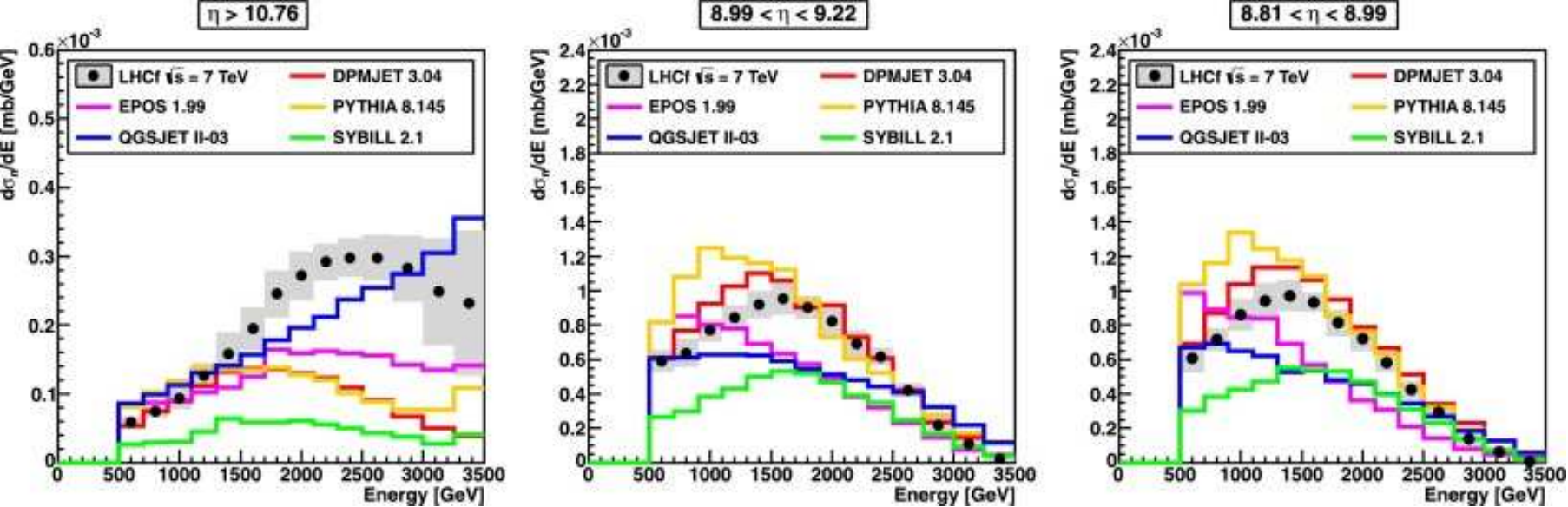}
\caption{A comparison of the LHCf neutron energy spectra measured in 7~TeV $pp$
  collisions \cite{LHCf-7TeV-n} with model predictions. The black markers and
  gray shaded areas show the corrected data and the systematic uncertainties,
  respectively.}
\label{LHCf}
\end{figure}
Figure~\ref{LHCf} shows the energy spectra of neutrons from the 7~TeV $pp$ collisions
\cite{LHCf-7TeV-n} where large deficiencies in all models used for modeling
cosmic rays (EPOS, QGSJET, DPMJET \cite{DPMJET} and SIBYLL \cite{SYBILL} and also
PYTHIA 8) are observed. The \pt\
spectra of $\pi^0$ produced in 7~TeV $pp$ collisions \cite{LHCf-7TeV-pi0} in a fine
$\eta$ binning are quite well described by EPOS, less well described by DPMJET.
Recently, photon energy spectra have been measured in 13~TeV $pp$ collisions
\cite{LHCf-13TeV-gamma} and again all models are observed to have difficulties in
describing the whole measured spectrum. 

\section{Identified particle spectra in $pp$, $pPb$ and $PbPb$}
\begin{figure}[h]
\begin{center}  
\includegraphics[height=.32\textheight, width=.331\textwidth]{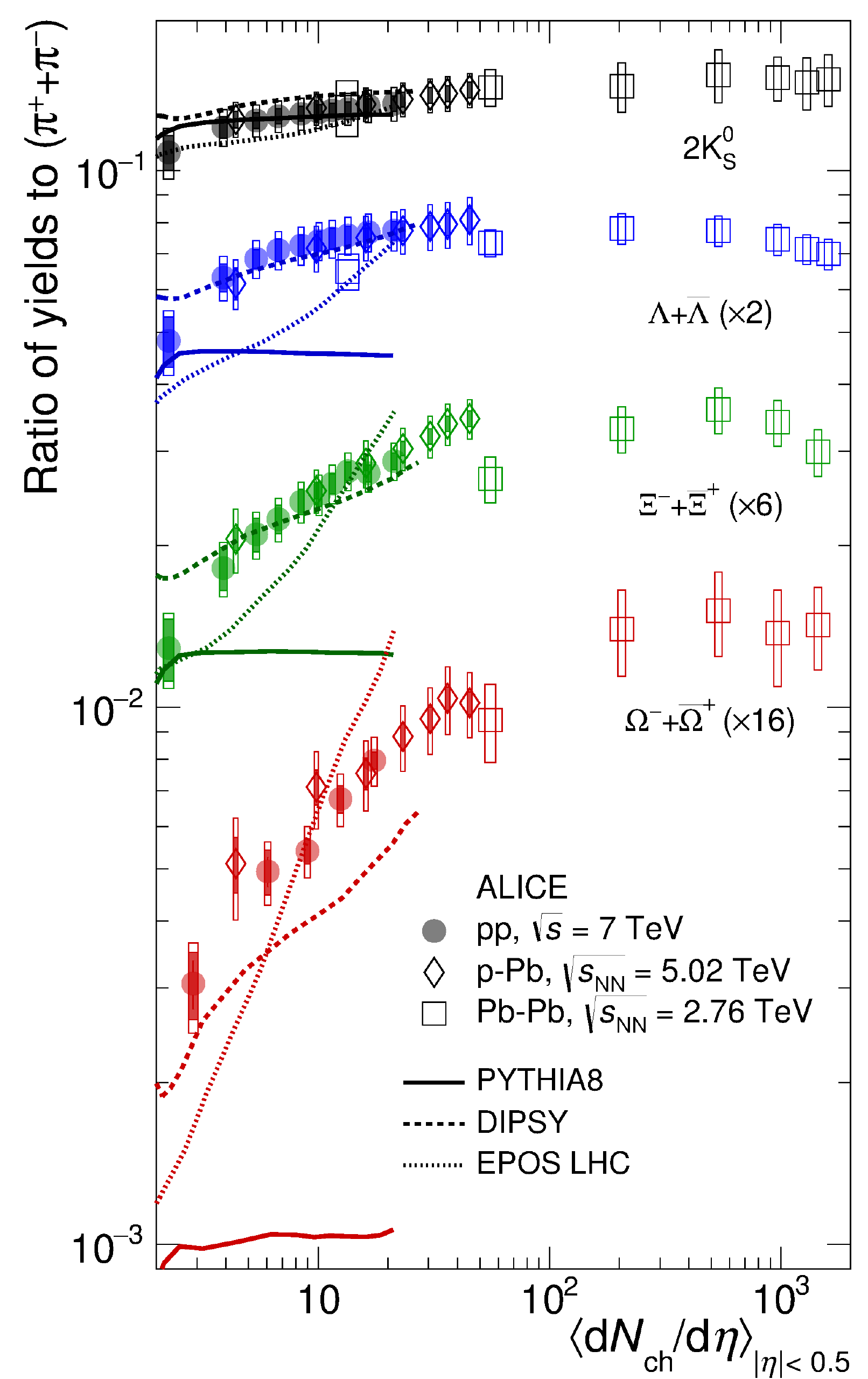}
  \hspace*{1.2cm}
  \includegraphics[height=.23\textheight, width=0.42\textwidth]{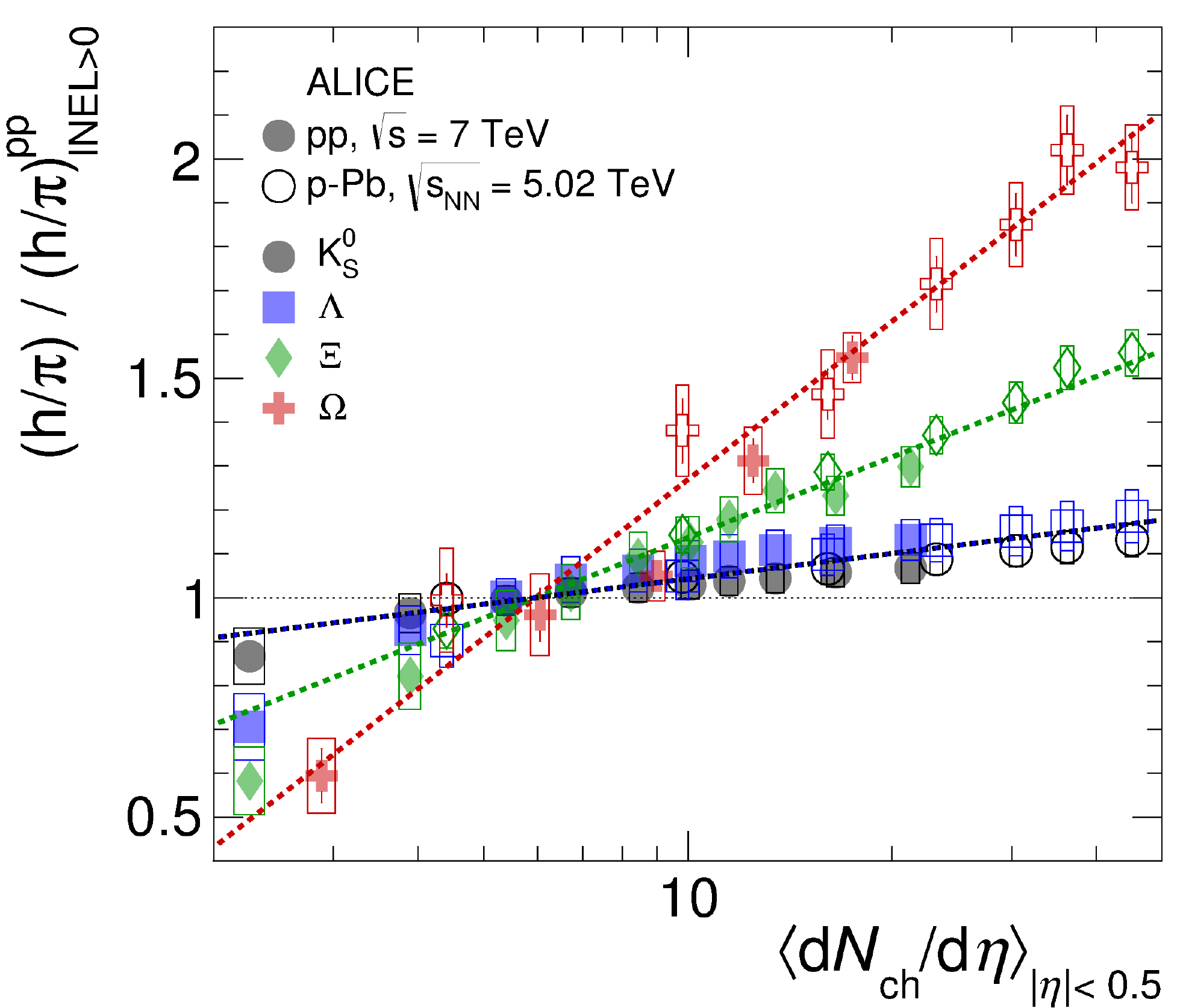}
\end{center}
\vspace*{-0.3cm}
  \caption{Left) The \pt\ integrated yield ratios to pions as a function of
    $\langle dn_{ch}/d\eta \rangle$ measured in $|\eta|<0.5$. The error bars show the
  statistical uncertainty, whereas the empty and dark-shaded boxes show the total
  systematic uncertainty and the contribution uncorrelated across multiplicity
  bins, respectively. The values are compared to calculations from MC models
  \cite{PYTHIA8,EPOS,DIPSY} and to previous results obtained by ALICE in $pPb$ and
  $PbPb$ collisions.
  %\cite{Abelev:2013haa,ABELEV:2013zaa,Adam:2015vsf}.
    Right) Particle yield ratios to pions normalized to the values measured in the
    inclusive $pp$ sample. The error bars show the statistical uncertainty. The
    common systematic uncertainties cancel in the double-ratio. The empty boxes
    represent the remaining uncorrelated uncertainties. The lines represent a
    simultaneous fit of the results with an empirical scaling formula.
    Plots taken from Ref.~\citenum{ALICE-id}.}
\label{ALICE-str}
\end{figure}

In order to identify particle species, each experiment has sophisticated
identification procedures usually based on the ionization energy loss, $dE/dx$,
or other techniques. In the ALICE analysis \cite{ALICE-id}, strange hadrons with
different strangeness content are identified and a ratio of multiplicity of
strange hadrons to pions is plotted in Fig.~\ref{ALICE-str} left as a function of
multiplicity density in the center of the detector for all three collision
systems: going from relatively low multiplicity densities occurring in $pp$
collisions over middle densities seen in $pPb$ collisions up to large densities
corresponding to $PbPb$ collisions. We observe that the integrated yields of
strange and multi-strange particles relative to pions, increase significantly
with the charged particle multiplicity and that the yield ratios measured in $pp$
collisions agree with $pPb$ measurements at similar multiplicities.
This is a first observation of a strangeness enhancement in high-multiplicity $pp$
collisions. In very high-multiplicity $pPb$ collisions, the strangeness production
reaches values similar to those observed in $PbPb$ collisions, where a quark-gluon
plasma (QGP) is formed. QGP is matter in a phase of deconfined quarks and gluons
which is created at sufficiently high temperatures and energy densities which is
usually reached in collisions of heavy ions with high energies. Strangeness
enhancement has been proposed as one of the main signatures of formation of
QGP \cite{strangeHI}. We see that none of the three models is able to describe the
data but DIPSY \cite{DIPSY}, a special model using color ropes, describes data
best. In the analysis \cite{ALICE-id}, the ratio of baryon to meson yields is also
studied and it is observed that none of the models is able to describe the data.
In Fig.~\ref{ALICE-str} right, the strangeness enhancement with respect to
inclusive samples is plotted: here we see a clear strangeness hierarchy, namely
the slope increasing with increasing strangeness content, and in addition, the
same hierarchy as measured already for $pPb$ data is observed. We conclude that
mass and multiplicity dependencies of the strangeness enhancement as well as of
the spectral shapes (not shown here, see \cite{ALICE-id}) remind the patterns
seen for $pPb$ and $PbPb$ collisions which can be understood assuming a
collective expansion of the system in the final state. 

Absolute yields and \pt\ spectra of identified particles in hadron-hadron
collision are usually used to improve the modeling of various key ingredients of
MC event generators, such as MPI, parton hadronization, and
final state effects such as parton correlations in color, \pt, spin, baryon and
strangeness number, and collective flow. Parton hadronization and final state
effects are mostly constrained from $e^+e^-$ data whose final states are dominated
by simple $q\bar{q}$ states, whereas low \pt\ hadrons at LHC come from
fragmentation of multiple gluons, so called minijets.  
This is also the reason why the production of baryons and strange hadrons in $pp$
collisions is not well reproduced by current generators, and hence makes a
good motivation for this study. In addition, identified particle spectra serve as
an important reference for high-energy HI studies. 
This CMS study \cite{CMS-13TeV-id} used data with negligible pile-up. Pions,
kaons and protons are identified using $dE/dx$, a specific ionization which works
very well for momenta lower than 1.2, 1.1 and 1.7~GeV for pions, kaons and
protons, respectively.
All particle species are limited to the region $|\eta| < 1.0$. And because
transverse momenta as low as 0.1~GeV are measured, special tracking algorithms
were used with high reconstruction efficiency and low background. These algorithms
feature special track seeding and cleaning, hit cluster shape filtering, modified
trajectory propagation and track quality requirements.
\begin{figure}[h]
  \begin{center}
  \includegraphics[height=.25\textheight, width=0.45\textwidth]{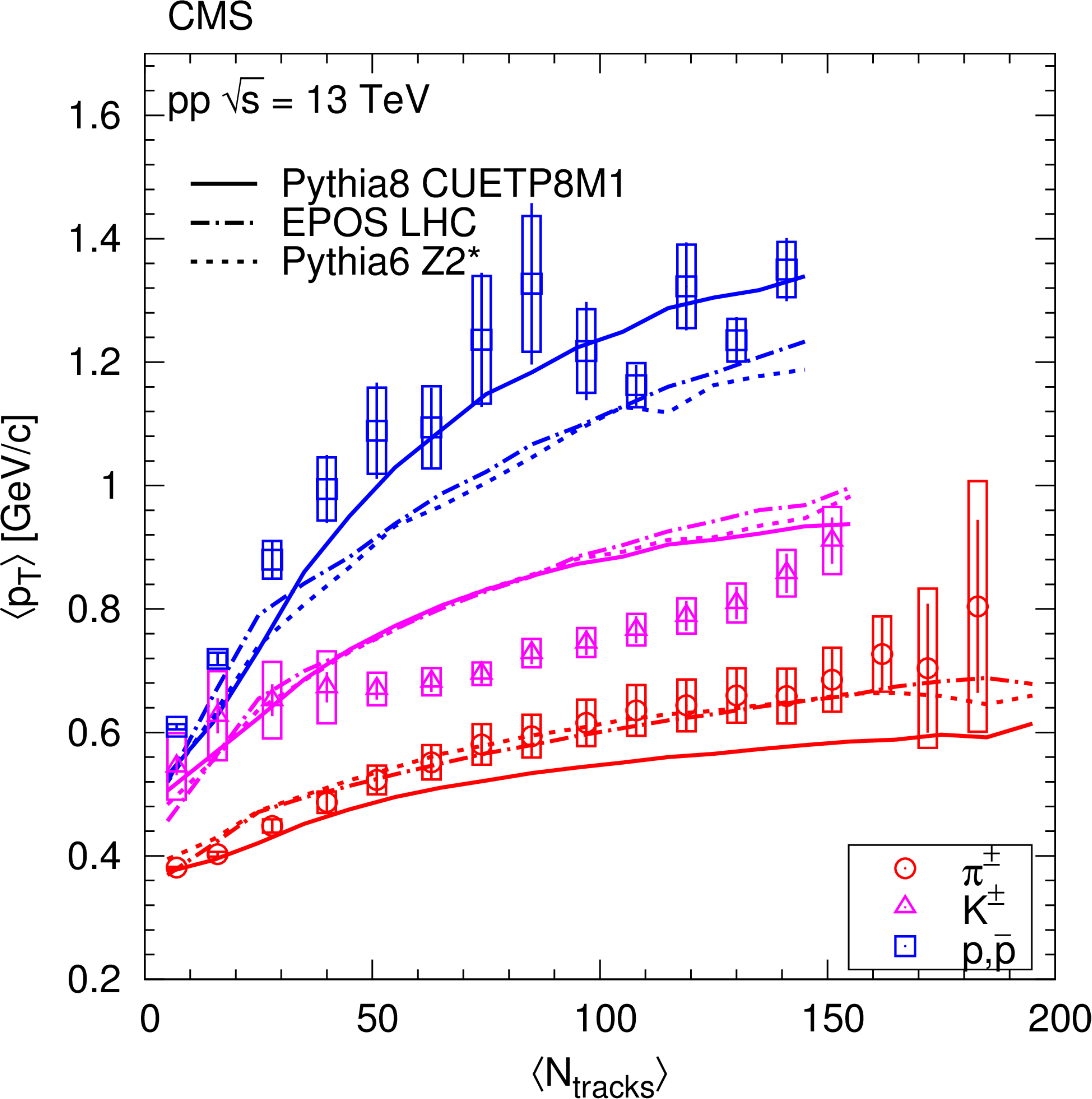}
  \includegraphics[height=.25\textheight, width=0.45\textwidth]{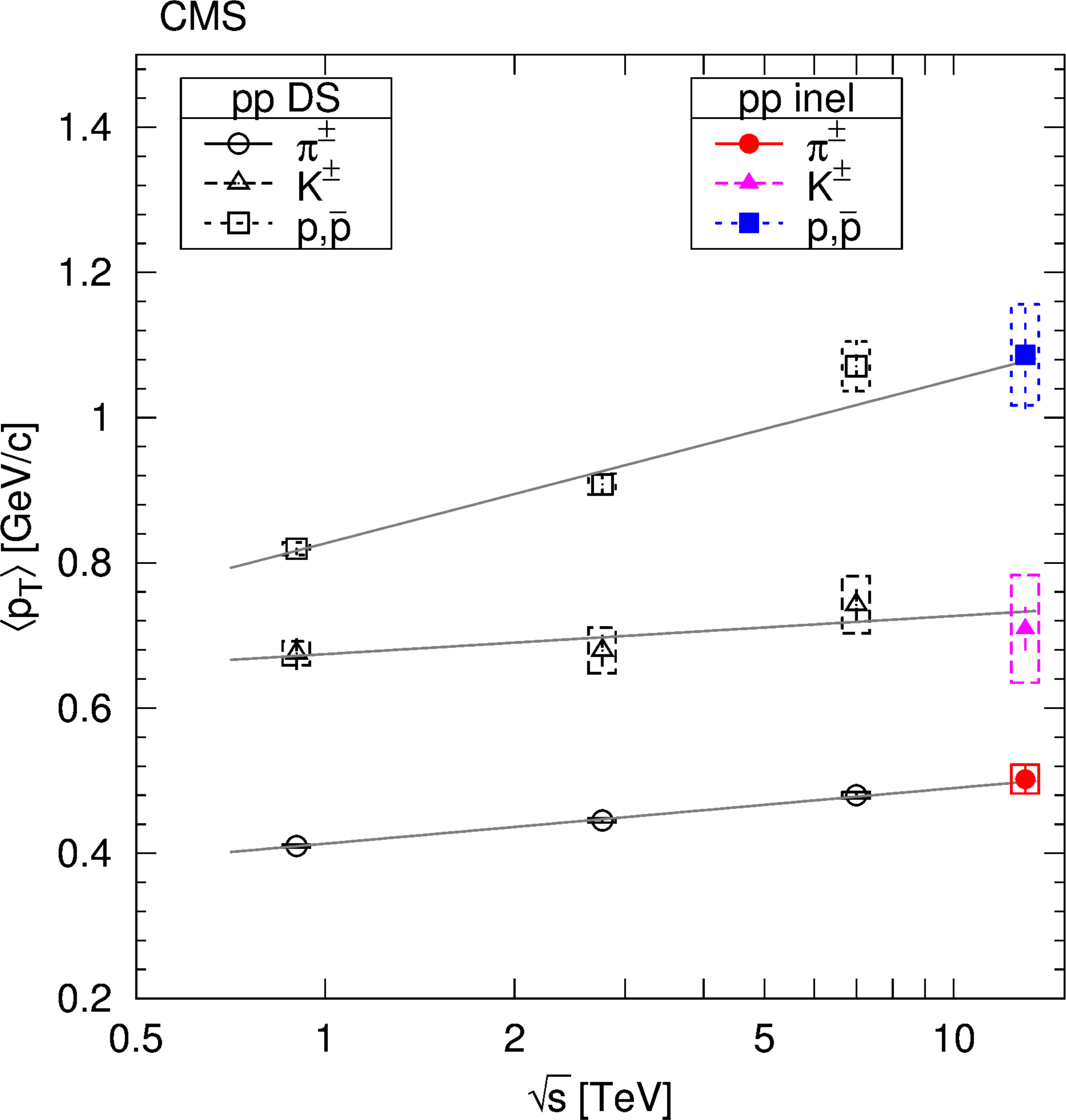}  
  \caption{Average \pt\ of charged-averaged pions, kaons and protons in the range
    $|\eta|< 1$ as functions: left) of the corrected track multiplicity for
    $|\eta|< 2.4$, computed assuming a Tsallis-Pareto distribution in the
    unmeasured range. Curves indicate predictions from PYTHIA~8, EPOS and
    PYTHIA~6; right) of \sq. The curves show linear fit in $\ln s$.
    Error bars indicate the uncorrelated combined uncertainties, while boxes show
    the uncorrelated systematic uncertainties. Plots taken from
    Ref.~\citenum{CMS-13TeV-id}.}
  \label{CMS-13TeV-idp}
  \end{center}
\end{figure}
Figure~\ref{CMS-13TeV-idp} left shows the average \pt\ as a
function of multiplicity in the event. We can notice that for kaons, all models
overshoot the data. While the low multiplicity region is well modeled by the
generators, the high multiplicity region needs some tuning of baryon and/or
strangeness production. Figure~\ref{CMS-13TeV-idp} right shows a rising evolution
of the average \pt\ with \sq. This
collision energy evolution of the average \pt\ provides useful information on the
so called saturation scale of the gluons in proton.

\section{Underlying event in $pp$}
The hard scattering is accompanied by interactions of a soft nature, namely
those coming from the rest of the proton-proton collision. They can come from the
initial state radiation (ISR), final state radiation (FSR), MPI and from color
reconnections (CR, namely from the QCD evolution of colour reconnections between
the hard scatter and beam remnants). A combination of contributions from all these
processes is called underlying event (UE) which is important to consider and
measure (or estimate) since: 
\begin{itemize}
\item These processes can not be completely described by perturbative QCD, and
  require phenomenological models, whose parameters are tuned by means of fits to
  data.
\item Final state (or its part) coming from UE can mimic a signal final state,
  for example the same-sign WW production from MPI can mimic final state of the
  same-sign dilepton SUSY searches.  
\item It can affect isolation criteria applied to photons and charged leptons.
\item It can affect the vertex reconstruction efficiency. For example the primary
  vertex in the process $H\rightarrow\gamma\gamma$ can be partly determined from
  the charged particles originating from UE.
\end{itemize}
An usual procedure of estimating the amount of UE is spatially dividing tracks in
each event according to their azimuthal angle to the Towards region (where the
highest \pt\ jet points), the Away region (where the second highest \pt\ jet
points) and then we have two Transverse regions where one of them contains
imprints of ISR and FSR and the other one has the least activity (so called
Transverse-min) --- this one is
believed to be the most sensitive to UE. The usual observables are average track
multiplicity per unit area and average scalar sum of track \pt\ per unit area. 
%The top-right plot shows the angular distribution of the \pt\ sum for two
%selections of the leading particle according to its \pt. The difference between
%these two selections illustrates the transition from relatively isotropic MB
%scattering to the emergence of hard partonic scattering structure and hence a
%dominant axis of energy flow. This event structure with least activity
%perpendicular to the leading-object axis, i.e. away from 0 and 180 degrees is seen
%for both selections but is much stronger for $p_{T,\mr lead} > 10$~GeV.

\begin{figure}[h]
\includegraphics[height=.21\textheight,width=0.31\textwidth]{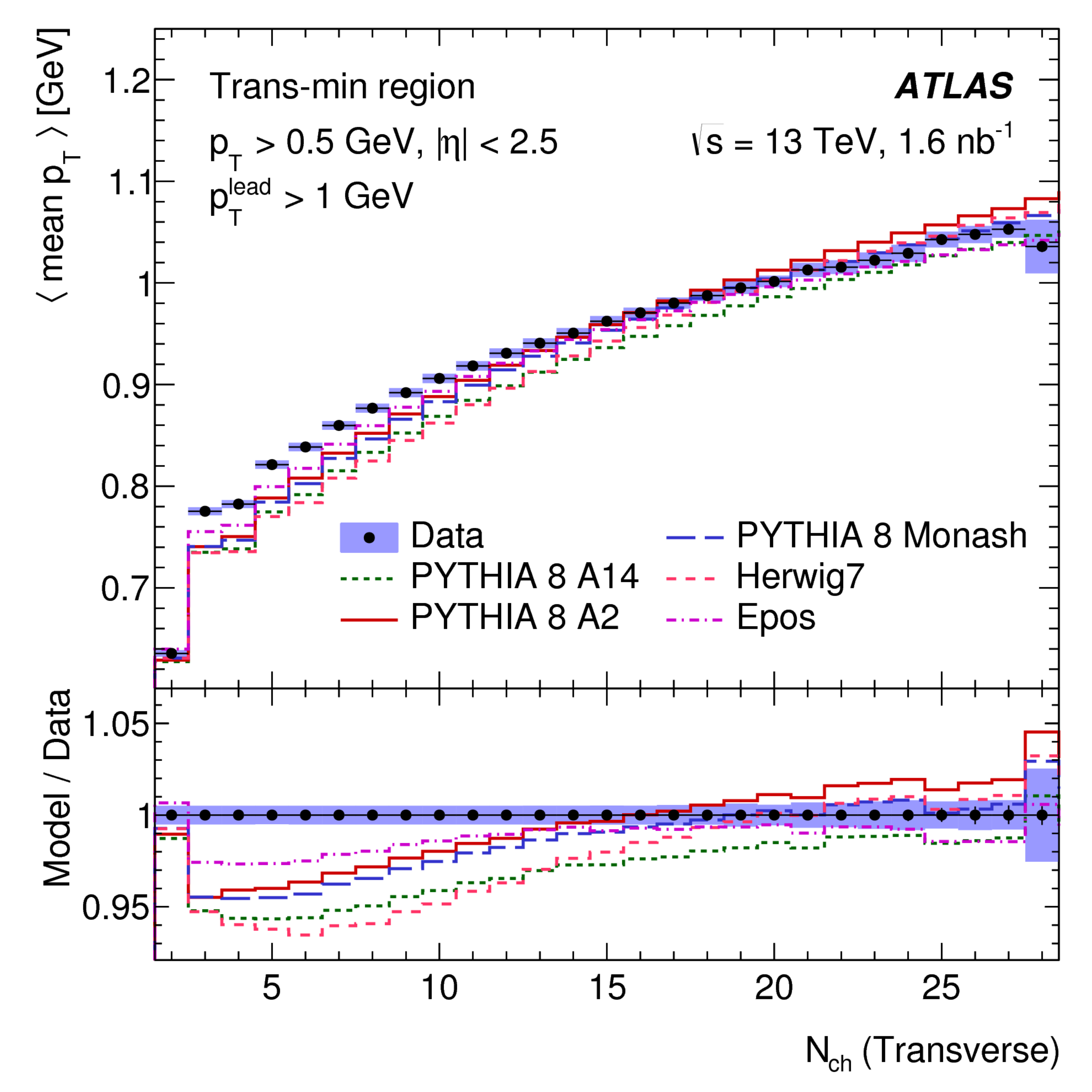}
\includegraphics[height=.21\textheight,width=0.33\textwidth]{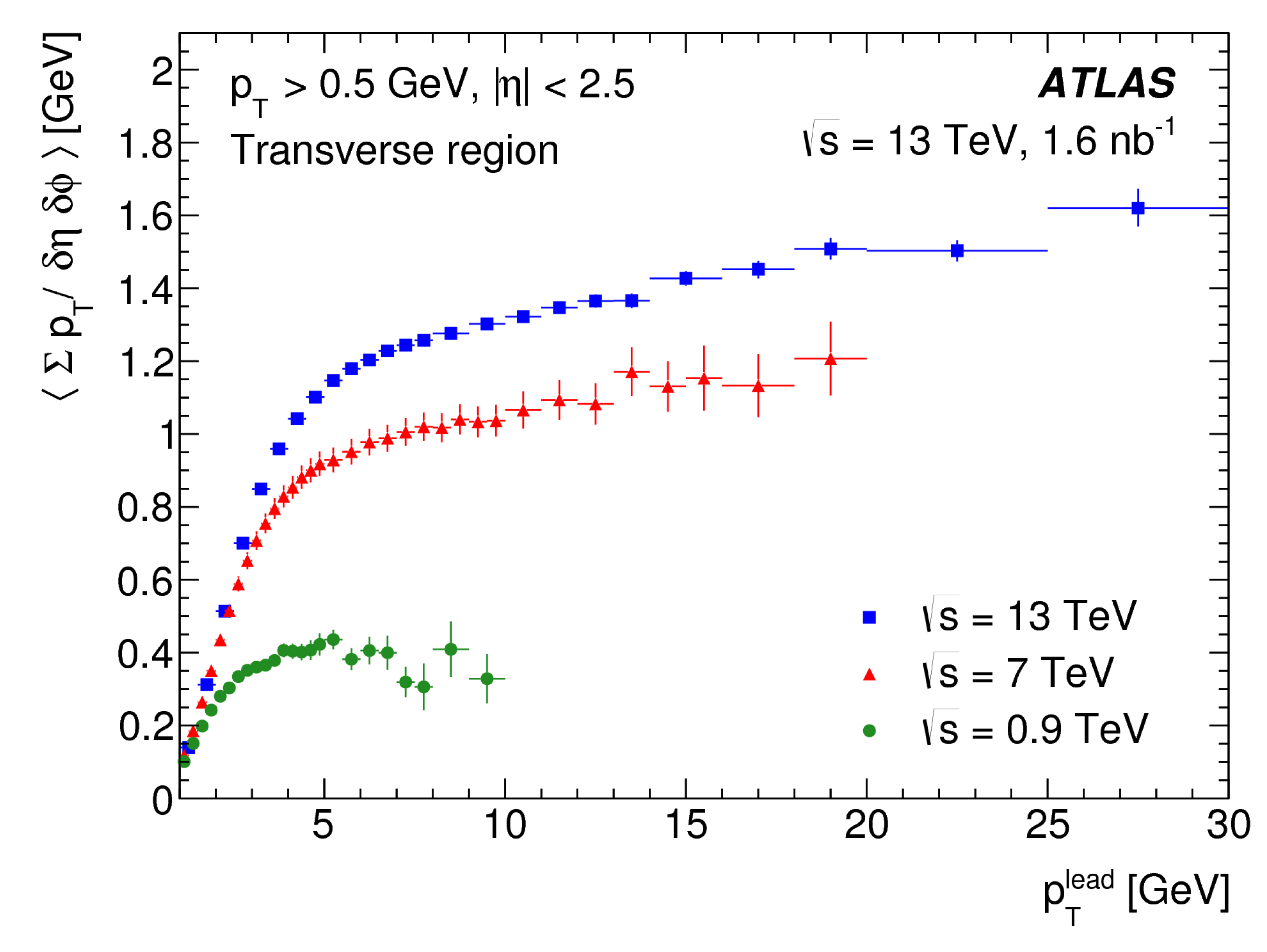}
\includegraphics[height=.22\textheight,width=0.345\textwidth]{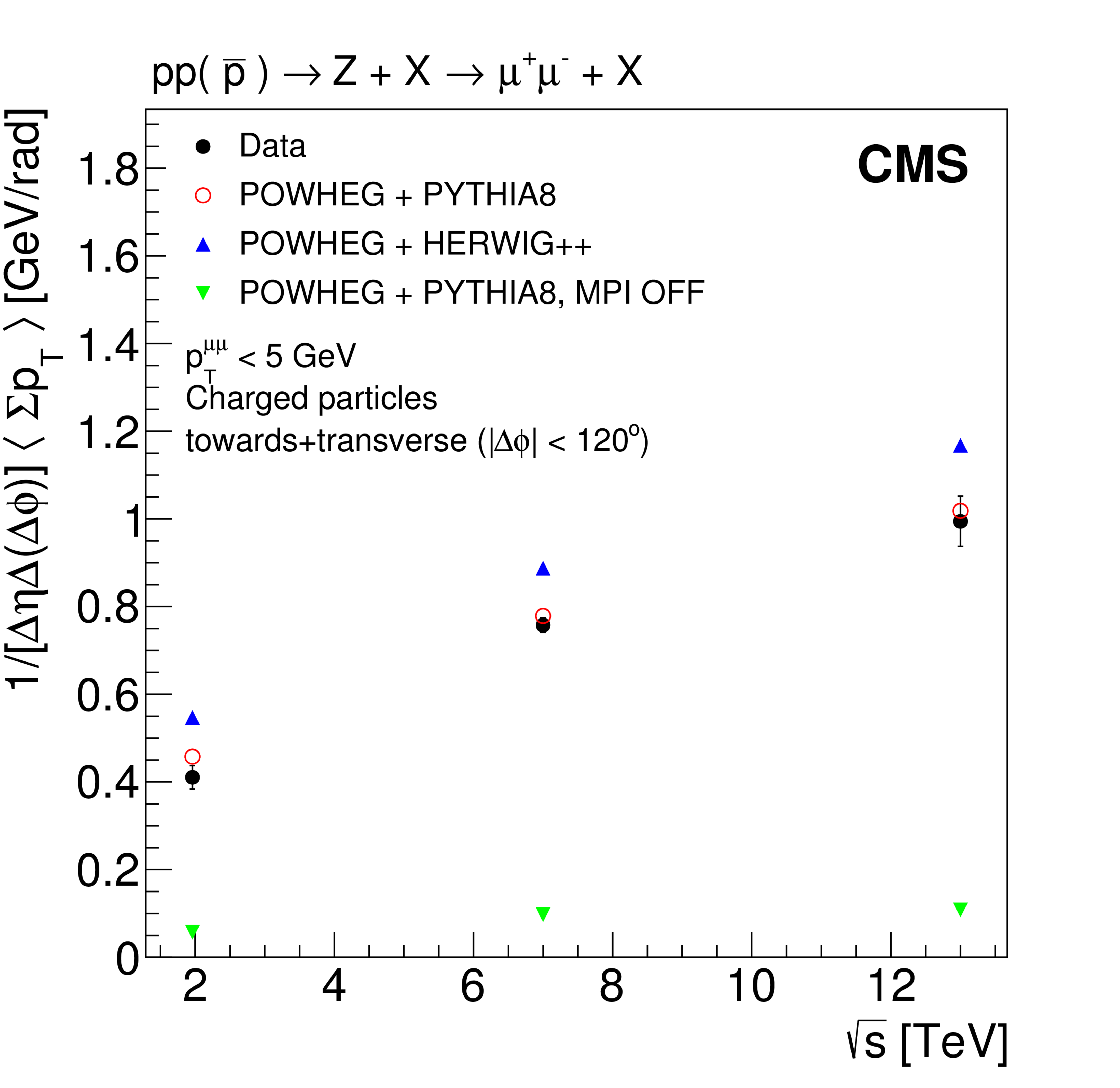}
  \caption{Left) Mean charged-particle average \pt\ as a function of
    charged-particle multiplicity in the transverse-min azimuthal region
    \cite{ATLAS-13TeV-UE}.
    The error bars on data points represent statistical uncertainty and the blue
    band the total combined statistical and systematic uncertainty. Middle) Mean
    sum \pt\ densities as a function of transverse momentum of the leading
    charged particle measured for \sq\ = 0.9, 7~TeV \cite{ATLAS-097TeV-UE} and
    13~TeV \cite{ATLAS-13TeV-UE}.
    %The fiducial acceptance definitions of the √s = 0.9 and 7TeV measurements did
    %not exclude charged strange baryons, but this effect is limited to a few
    %percent at most.
    Right) Sum \pt\ density, with $p_{\mu\mu} < 5$~GeV as a function of \sq\ for
    data \cite{CMS-13TeV-UE} and predictions from simulations by POWHEG+PYTHIA~8
    and POWHEG+HERWIG++.
    The predictions of POWHEG+PYTHIA~8 without MPI are also shown. Error bars
represent the statistical and systematic uncertainties added in quadrature.}
\label{13TeV-UE}
\end{figure}

Figures~\ref{13TeV-UE} left and middle concentrate on the UE-dominated region
studied in ATLAS events containing at least one charged particle with
$p_T > 1$~GeV\cite{ATLAS-13TeV-UE}.
The left plot shows the multiplicity evolution of the average \pt, it can also be
seen as a
correlation between two ``soft'' properties. It shows a balance between \pt\ sum
and multiplicity. This balance is affected in some models by CR which typically
increases the \pt\ per particle. The description by all models is within 5\%.
EPOS gives in general best description at low \pt\ of the leading particle, but is
the worst when the \pt\ of the leading particle is above 10~GeV (not shown here). 
The middle plot shows the evolution of the \pt\ sum density with \pt\ of the
leading particle, for three collision energies. The initial rapid rise up to
$p_T^{\mr lead}\approx 5$~GeV after which the density stabilizes around one 
charged particle or 1~GeV per unit $\phi-\eta$ area is known as the ``pedestal
effect''. It reflects a reduction of the $pp$ impact parameter with increasing
$p_T^{\mr lead}$ and hence the transition between the minimum bias and hard
scattering regimes.
The right plot shows a similar trend, namely the evolution of \pt\ average
per event with increasing collision energy. It is a CMS \cite{CMS-13TeV-UE}
analysis based on Drell-Yan events selected using a detection of a dimuon pair.
From a comparison to various models and mainly to PYTHIA with and without MPI, we
can clearly see the importance of MPI in modeling the UE activity. 

\section{2-particle azimuthal correlations in $pp$, $pPb$ and $PbPb$}
Soft processes are also in the heart of heavy-ion collisions \cite{M+R}. In the
HI collisions at LHC, it is believed that quark-gluon plasma is created. The
conditions for the phase transition
are created, the system gets close to the thermal equilibrium and expands
collectively. The expansion means that the matter cools down and hadrons are
formed. The aim is to measure macroscopic properties of the QGP and study its
microscopic laws.
%But there are new challenges and opportunities in the recent results in $pp$ and
%$pPb$. 
Usually $PbPb$ collisions serve to create and study QGP, the $pPb$ collisions serve
as a sort of control experiment (cold nuclear matter effects (e.g. modifications
to PDF)) and the $pp$ collisions serve as reference data. 
But recently, striking similarities were observed between these three collision
systems. Phenomena considered fundamental for QGP are now seen also in $pPb$ and
even $pp$. These were discovered in high-multiplicity events but they may be
relevant also for minimum bias events which would, in turn, have important
consequences for
all hadronic collisions. We speak especially about the ridge observation. The
ridge was first observed in central HI collisions as 2-particle long-range
correlations ($\Delta\eta > 2$) on a near side ($\Delta\phi \approx 0$) and it is
believed to be a result of collective
hydrodynamic expansion of hot and dense nuclear matter created in the overlap
region. The 2-particle correlations exhibit also the so-called away-side peak
coming from the other lower-energy jet. The ridge is usually described by Fourier
decomposition using a term $\cos(n\Delta\phi)v_n$ where $v_n$ is a
single-particle anisotropy harmonics. Unexpectedly ridge structures were also
measured in $pPb$ and even in $pp$ at high multiplicities.
%The four pictures just illustrate the size of the ridge when going from HI
%collision through $pPb$ collision down to high-multiplicity events in $pp$ collision
%where you can observe a small but visible ridge. The low-multiplicity $pp$ events
%do not have any ridge as shown on the rightmost picture.
Currently the origin of the ridges in these small collision systems is lively
debated. Is it hydrodynamics like in QGP which would mean a final state effect?
Or is it due to initial state fluctuations (as embedded in Color Glass Condensate
(CGC) model \cite{CGC} using gluon saturation)? Or perhaps these long range
correlations come from hadronization described by ropes \cite{DIPSY}? Or from
collisions of thin flux tubes \cite{Bjorken_ridge}? Definitely the ridge is a
testing ground to study complementarity between dynamical and hydrodynamical
models.

Particle correlations are a very powerful tool to study properties of multihadron
production in general. Several production mechanisms can be studied
simultaneously. The baseline mechanism underlying all correlations is a global
conservation of momentum and energy as well strangeness, baryon number and
electric charge. Other phenomena, including mini-jets, elliptic flow,
Bose-Einstein correlations (BEC) and resonance decays are source of additional
correlations and all those sum up.

By studying the $\Delta\phi$ projections of 2-particle correlations around the
near-side and away-side peak, it was observed e.g by LHCb \cite{LHCb-5TeV-dphi}
using 5.02~TeV $pPb$ data and by ALICE \cite{ALICE-5TeV-dphi} using 2.76~TeV $PbPb$
data that both peaks increase with multiplicity and that the size of the
near-side peak is maximal for particle with $1<p_T<2$~GeV. 
\begin{figure}[h]
  \includegraphics[height=.24\textheight,width=0.48\textwidth]{plots/CMS-13TeV-corrnch.pdf}
  \includegraphics[height=.24\textheight,width=0.48\textwidth]{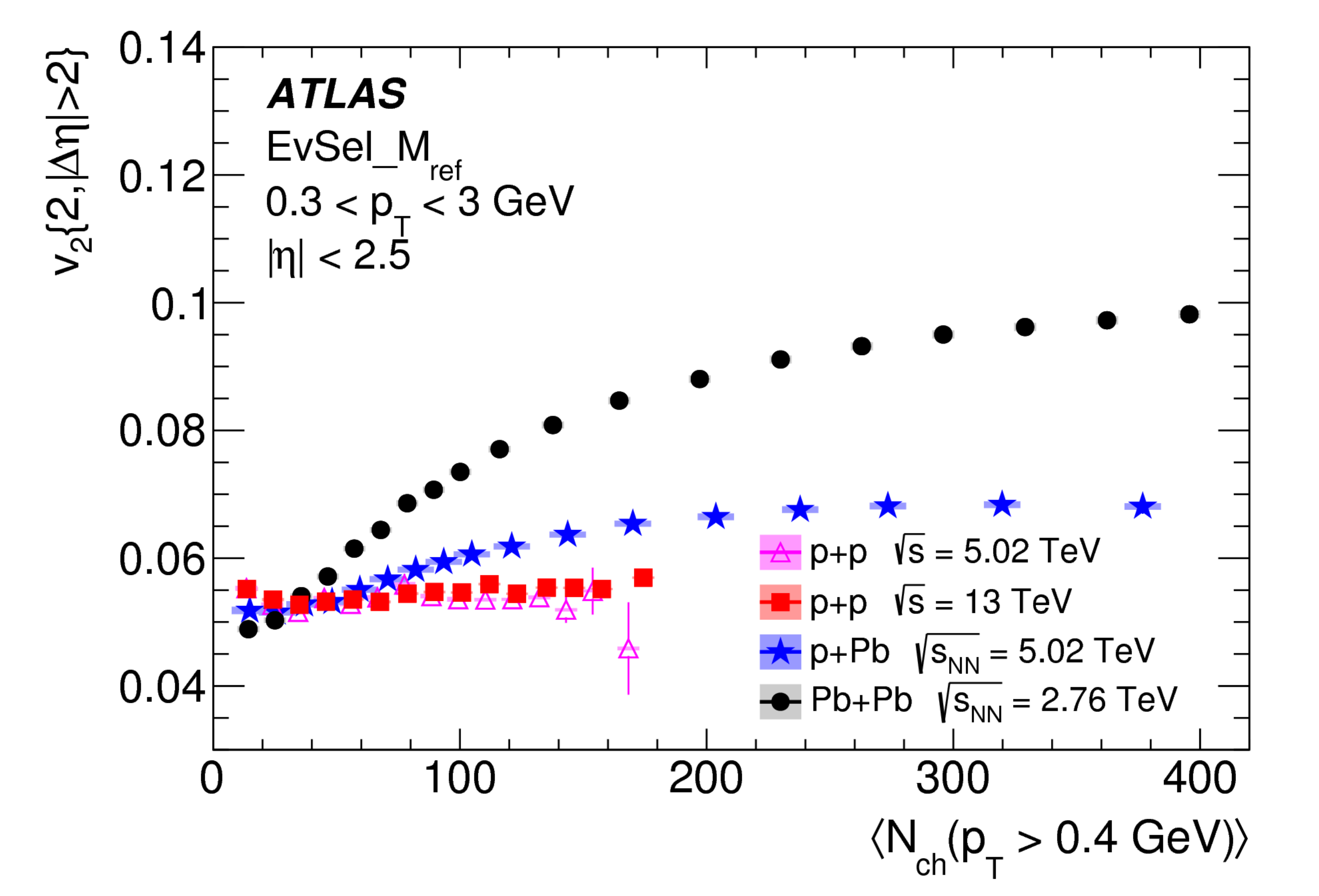}  
  \caption{Left) Associated yield of long-range near-side two-particle
    correlations for 1 $< p_T <$ 2~GeV in $pp$ collisions at \sq\ = 13 and 7~TeV,
    $pPb$ collisions at \sq\ = 5.02~TeV, and $PbPb$ collisions at \sq\ = 2.76~TeV.
    The error bars correspond to the statistical uncertainties, while the shaded
    areas denote the systematic uncertainties. Plot taken from
    Ref.~\citenum{CMS-corrnch}.
    Right) Comparison of $v_2\{2,|\Delta\eta|>2\}$ as a function of
    $\langle N_{ch}\rangle(p_T > 0.4$~GeV)⟩ for $pp$ collisions at \sq\ = 5.02 and
  13~TeV, p+Pb collisions at \sq\ = 5.02~TeV and low-multiplicity Pb+Pb
  collisions at \sq\ = 2.76~TeV, and for reference particles with 0.3
  $< p_T <$~3~GeV. The error bars and shaded boxes denote statistical and
  systematic uncertainties, respectively. Plot taken from
  Ref.~\citenum{ATLAS-v2nch}.}
\label{2pc}
\end{figure}
To study the long-range correlations, however, we have to subtract 2-particle
correlations coming from the so called ``non-flow'' which includes resonance
decays and dijets. There are several methods to do that, one class of methods
tries to subtract the non-flow using low-multiplicity events. It also turns out
that the extraction of collective flow in $pp$ collisions strongly depends on the
event selection and also on the purity of the non-flow extraction as documented
e.g. in the analysis~\cite{ATLAS-3subev}.

Figure~\ref{2pc} left shows an almost linear increase of 2-particle correlations
with event multiplicity for all three collision systems measured by CMS
\cite{CMS-corrnch}. The size of these
correlations is biggest for HI collisions, while it is smallest for $pp$ collisions.
Figure~\ref{2pc} right then shows a measurement by ATLAS \cite{ATLAS-v2nch} of the
$v_2$ quantity, the elliptic flow
harmonics, again for all three collision systems now as a function of
multiplicity. The measurement shows that $v_2$ from $pPb$ is smaller than that
extracted from the $PbPb$ collisions but originally much smaller values from $pPb$
collisions were expected reflecting the big difference in sizes of the $PbPb$ and
$pPb$ systems.

\section{Multi-particle azimuthal correlations in $pp$, $pPb$ and $PbPb$}
In the light of difficulties with the residual non-flow which the 2-particle
correlations suffer from, methods based on multi-particle correlations started to
be more widely used. It was proved that multi-particle correlations are
more robust with respect to the non-flow but it is clear that they are also more
statistically demanding. The method often used is to build cumulants $c_n\{2k\}$
(of the $n$-th order based on $2k$-particle correlations) and calculate flow harmonics $v_n\{2k\}$ from them. A novel method, the
three-subevent method, has recently been proposed by ATLAS \cite{ATLAS-3subev}
which turns out to be quite effective in reducing the non-flow and moreover it
provides negative values of cumulants $c_2\{4\}$, $c_2$ calculated from 4-particle
correlations, which is required to get positive $v_2\{4\}$. 
\begin{figure}[h]
  \includegraphics[height=.23\textheight,width=0.98\textwidth]{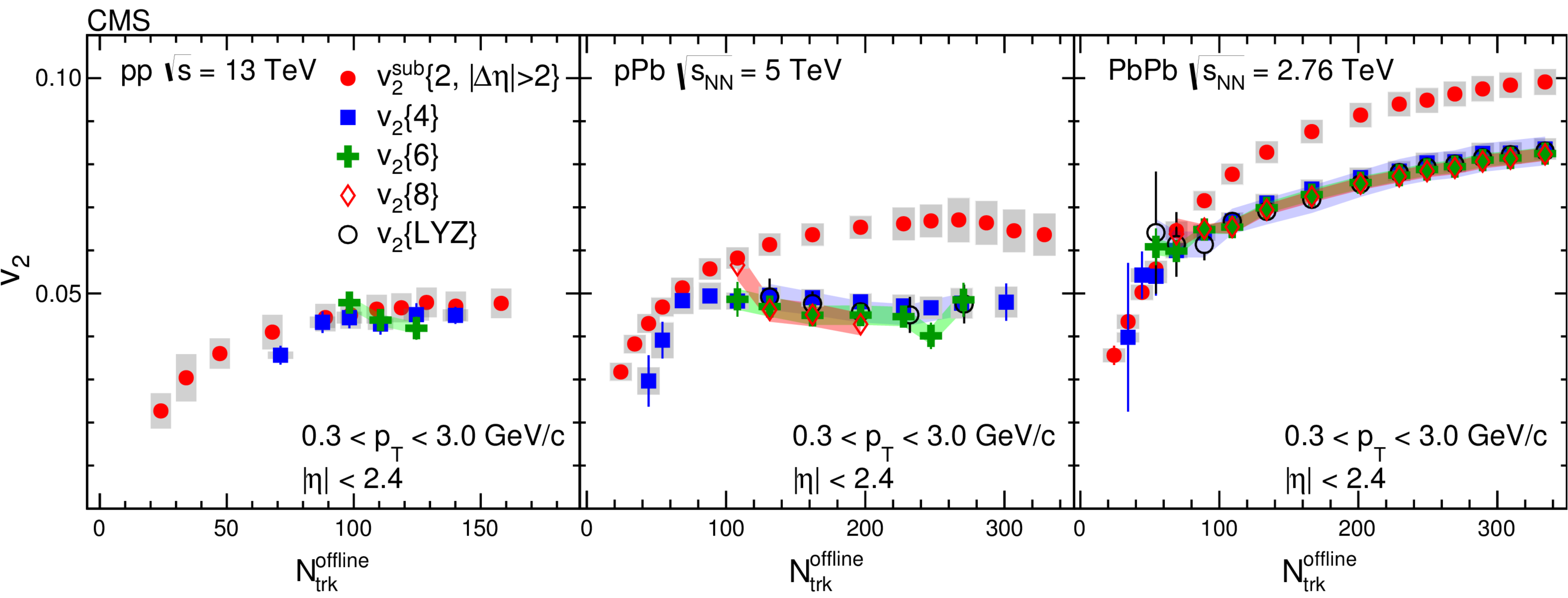}  
  \caption{Left) The $v_2$ values calculated from 2-, 4-, 6- and 8-particle
    correlations as functions of multiplicity of charged particles, averaged over
    0.3 $<p_T<$ 3.0~GeV and $|\eta|< 2.4$, in $pp$ collisions at \sq\ = 13~TeV
    (left), in $pPb$ collisions at \sq\ = 5.02~TeV \cite{CMS-mpc} (middle) and in
    $PbPb$ collisions at \sq\ = 2.76~TeV \cite{CMS-mpc} (right). The error bars
    correspond to the statistical uncertainties, while the shaded areas denote
    the systematic uncertainties. Plot taken from Ref.~\citenum{CMS-multipc}.}
\label{multipc}
\end{figure}
Figure~\ref{multipc} from the multi-particle correlation study by CMS
\cite{CMS-multipc} shows $v_2$ coefficients calculated using 2-, 3-, 4-, 6-
and even 8-particle correlations as functions of multiplicity for all three
collision systems. First the results say that $v_2$ from 4-particle correlations
are smaller than $v_2$ from 2-particle correlations in $pPb$ and $PbPb$ collisions -
that was expected for long-range correlations. We however observe smaller
$v_2\{4\}$ than $v_2\{2\}$ values for the $pp$ system (even larger differences are
reported using the three-subevent method in the analysis\cite{ATLAS-3subev}) and
a similarity between $v_2\{4\}$ and $v_2\{6\}$ values in all three collision
systems. All these findings suggest again that some collective effects are
occurring even in $pp$ collisions. 

\section{Angular correlations of identified particles in $pp$}
The 2-particle correlations can also be studied with identified particles.
By choosing specific particle types, we select a specific combination of quantum
numbers (strangeness, baryon number) that may manifest in the measured
correlations. The correlations should also be sensitive to the details of particle
production, including the parton fragmentation. In this ALICE study
\cite{ALICE-corrid} the near-side peak structure is studied. It is a combination
of at least three effects: i) fragmentation of hard-scattered
partons, ii) resonance decays and iii) femtoscopic correlations (BEC for identical
bosons, Fermi-Dirac anticorrelations for identical fermions, Coulomb and
strong final-state interactions). For pairs of same mesons, the near-side peak
comes from the minijet mechanism and BEC. The correlations of
particle-antiparticle pairs also include
a minijet like structure on the near-side as well as on the away-side. For pairs
of non-identical particles Bose-Einstein and Fermi-Dirac effects are not present,
however, resonances play a significant role. In contrast to same-sign meson
correlations, the baryon-baryon (or antibaryon-antibaryon) distributions for
identical proton and lambda baryon pairs show a qualitatively different effect,
namely a near-side depression instead of the peak. The correlations for same-sign
pions and kaons as well as same-sign protons and $\Lambda$ baryons are shown in
Fig.~10, projected onto the $\Delta\phi$ axis and
compared with several MC generators.
\begin{figure}[h]
  \begin{center}
\includegraphics[height=.5\textheight,width=0.85\textwidth]{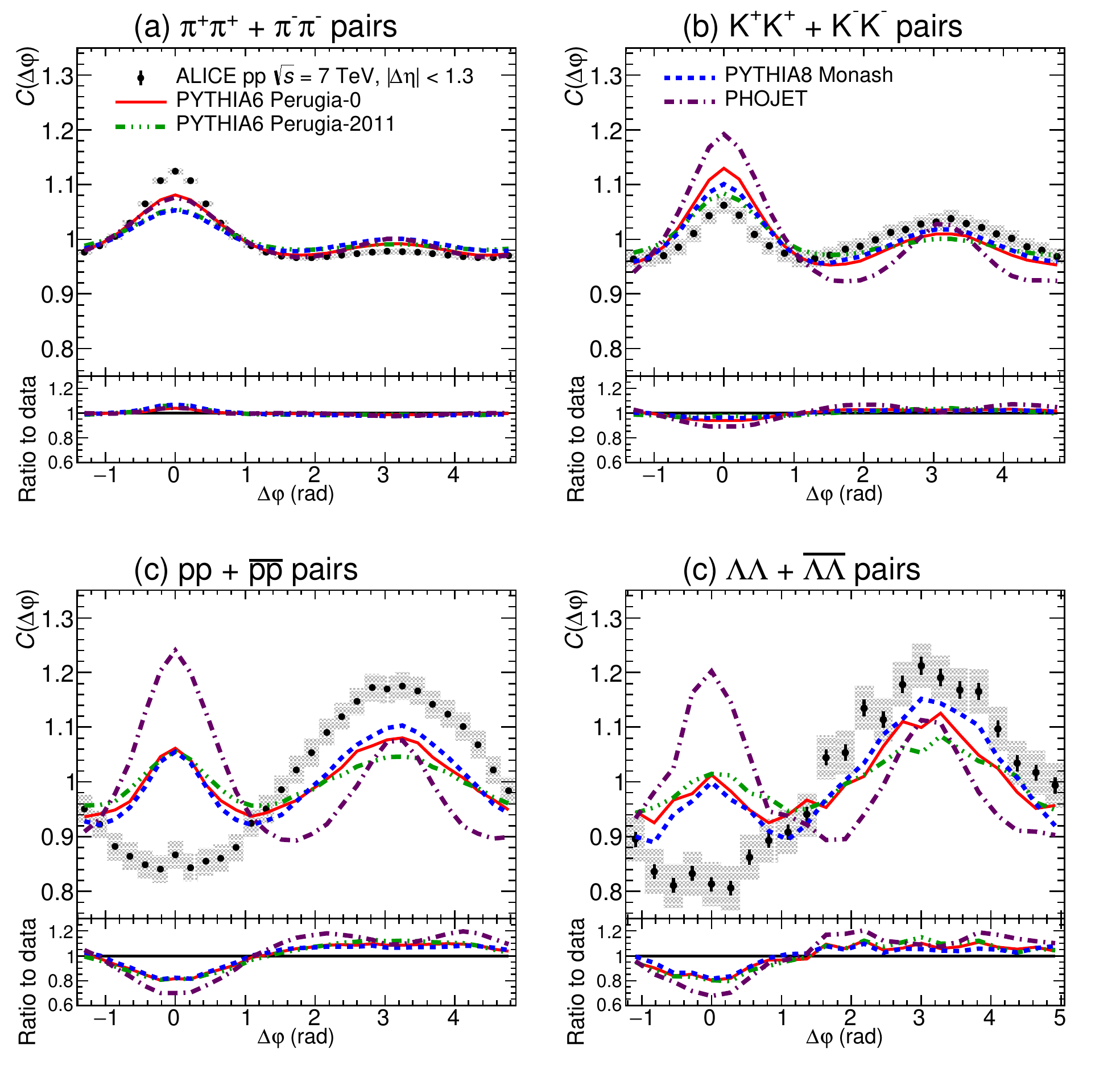}    
  \caption{The $\Delta\eta$ integrated projections of correlation functions for
    combined pairs of (a) $\pi^+\pi^++\pi^-\pi^-$, (b) $K^+K^++K^-K^-$, (c)
    $pp+\bar{p}\bar{p}$ and (d) $\Lambda\Lambda+\bar{\Lambda}\bar{\Lambda}$, obtained from
    ALICE data and four Monte Carlo models (PYTHIA~6 Perugia-0, PYTHIA~6
    Perugia-2011, PYTHIA~8 Monash, PHOJET). Bottom panels show ratios of MC models
    to ALICE data. Statistical (bars) and systematic (boxes) uncertainties are
    plotted. Plot taken from Ref.~\citenum{ALICE-corrid}.}
  \end{center}
\label{ALICE-2pcid}
\end{figure}
While for mesons (a-b), the description of the apparent near-side and small away
side peaks is reasonable, all event generators fail by giving positive
correlations for protons and $\Lambda$ baryons (c-d), where data show a
significant depression.
It should be noted that these generators conserve the
local baryon number and do not include quantum statistical effects such as
Fermi-Dirac anticorrelations. Several sources of this depression
were studied in this analysis and it was shown that neither Fermi-Dirac
anticorrelations nor strong final state effects nor local baryon number
conservation can explain this depression. So we conclude that something essential
is missing in the string fragmentation.

\section{Bose-Einstein correlations in $pp$, $pPb$ and $PbPb$}
ATLAS has recently published a BEC study with 7~TeV data based on particles with
\pt\ as low as 100~MeV \cite{ATLAS-7TeV-pp-BEC}. BE correlations are defined
using $c_2$, the 2-particle cumulants for identical particles, and are usually
plotted as a ratio of same-sign to opposite-sign $c_2$ cumulants as a
function of $Q$ which is the momentum difference of the particles in a pair.
This $c_2(Q)$ dependence is then fitted using the function
$C_2 = [1 + \Omega(\lambda, R)](1 + \epsilon Q)$ where $\lambda$ represents a
correlation strength and $R$ represents a size of the correlation source.
\begin{figure}[h]
  \begin{center}
  \includegraphics[height=.24\textheight,width=0.45\textwidth]{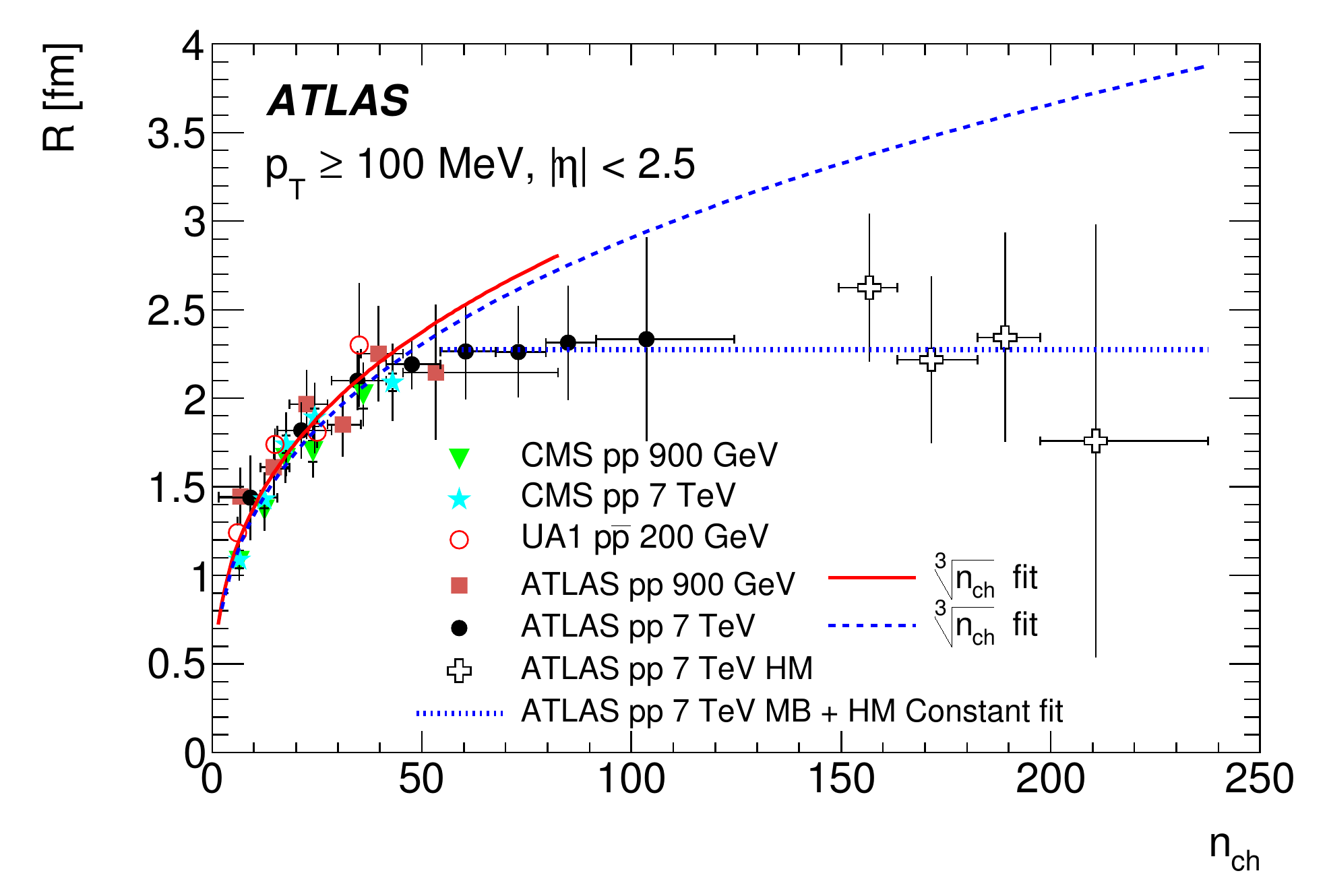} 
\includegraphics[height=.22\textheight,width=0.37\textwidth]{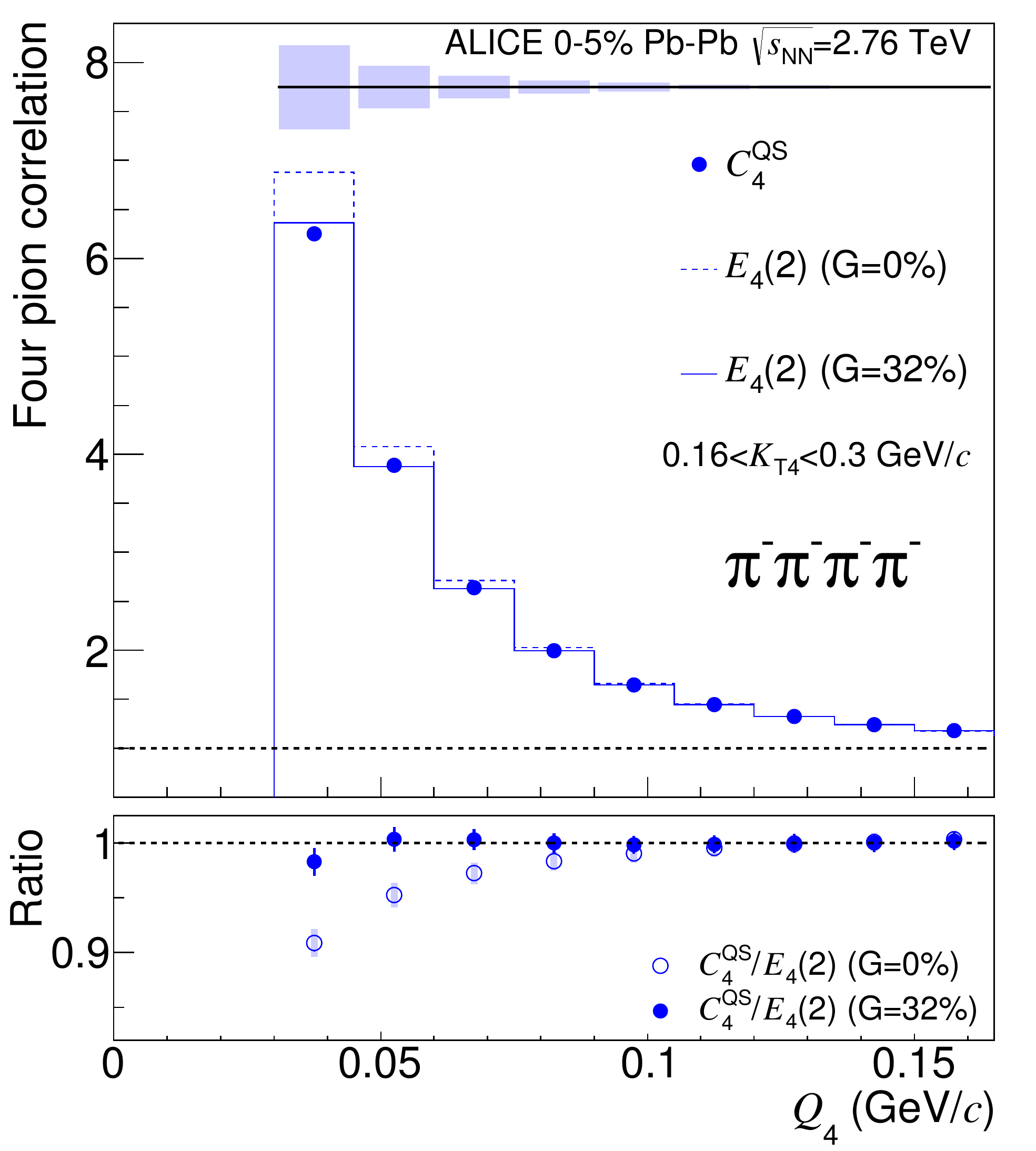}
  \caption{Left) Multiplicity $n_{\mr ch}$ dependence of the parameter $R$ obtained
    from the exponential fit to the two-particle double-ratio correlation
    functions $R_2(Q)$ at \sq\ = 0.9 and 7~TeV, compared to the equivalent
    measurements of the CMS \cite{CMS-BEC} and UA1 \cite{UA1-BEC} experiments.
    The solid and dashed curves are the results of the $\sqrt[3]{n_{\mr ch}}$ for
    $n_{\mr ch} <$ 55 fits. The dotted line is a result of a constant fit to
    minimum-bias and high-multiplicity events data at 7~TeV for $n_{\mr ch} \geq$
    55. The error bars represent the quadratic sum of the statistical and
    systematic uncertainties. Plot taken from Ref.~\citenum{ATLAS-7TeV-pp-BEC}.
    Right) Same-charge four-pion full correlations versus $Q_4$ (see the text).
    Measured (points) and expected (histograms) correlations of the first type
    are shown. Dashed and solid block histograms show the expected correlations
    based on G=0 and G=32\% fraction of coherent correlations, respectively.
    Systematic uncertainties are shown at the top. The bottom panel shows the
    ratio of measured to the expected correlations. The systematic uncertainties
    on the ratio are shown with a shaded blue band (G=0) and with a thick blue
    line (G=32\%). Plot taken from Ref.~\citenum{ALICE-pion-BEC}.}
  \label{BEC}
  \end{center}
\end{figure}
The size of the source has been measured as a function of event multiplicity and
particle \pt. The multiplicity dependence represented in Fig.~\ref{BEC} left shows
an interesting feature, namely a
rise and then a sudden stop and a saturation from multiplicities of about 50. This
saturation at high multiplicities is seen for the first time at all. The \pt\
dependence of $R$ was measured to decrease for all multiplicity classes. 
A similar tendency was measured in another ATLAS analysis of $pPb$ data
\cite{ATLAS-pPb-BEC}. Also in the LHCb study \cite{LHCb-BEC}, $R$ was measured to
increase with multiplicity, while $\lambda$ to decrease with multiplicity, so in
accordance with observations made by the other LHC experiments, we can conclude
that larger sources are more coherent.

Multi-pion BE correlations have also been measured in all three collision systems
by ALICE \cite{ALICE-pion-BEC}. In Fig.~\ref{BEC} right we can see the 4-pion
correlations as a function of $Q$ between these pions and the ratio of these
measured 4-particle correlations to the expected ones from 2-pion correlations.
While for $pp$ and $pPb$ collisions, no suppression is seen, in $PbPb$, a clear
suppression of both the 4-pion and 3-pion (not shown here) correlations is
observed. Interestingly, if a 32\%-fraction of coherent correlations is assumed,
the suppression is explained for the 4-pion correlation, while it does not help
to the 3-pion correlations.

\section{Hadronic chains}
The origin of the enhanced production of pairs of identical particles is usually
studied using BEC, believed to come from an incoherent particle production (see
the previous section). An alternative approach to BEC is a causality-respecting
model of quantized fragmentation of a 3D QCD string as a consequence of coherent
hadron emission. This approach is based on studying hadronic chains \cite{Sarka}
using helical strings \cite{helix} and utilized in a recent ATLAS analysis
\cite{ATLAS-hadchains2}. It studies hadronization effects using the same ATLAS
data as in the BEC study \cite{ATLAS-7TeV-pp-BEC}, namely again same-sign and
opposite-sign identical particle pairs. In the Lund hadronization, used in PYTHIA,
there is a randomly broken 1D string and no cross-talk between break-up vertices.
In the model of quantized helical (3D) string, one tries to make use of causality
(in other words the cross-talk between break-up vertices) which gives two
parameters: $\kappa R$ and $\Delta\phi$. This model predicts that the hadron
spectra follow a simple quantized pattern: $m_T = n\kappa R\Delta\phi$, so that
$\kappa R$ and $\Delta\phi$ can be fixed using known masses of pseudoscalar
mesons.
Then one can predict momentum difference $Q$ for pairs of ground-state hadrons
for various pair rank differences, $r$. Adjacent (opposite-sign) pions are then
predicted to be produced with $p_T$ difference of 266~MeV, while the like-sign
pion pairs with rank difference 2 should have 91~MeV. 
\begin{figure}[h]
  \begin{center}
  \includegraphics[height=.15\textheight,width=0.45\textwidth]{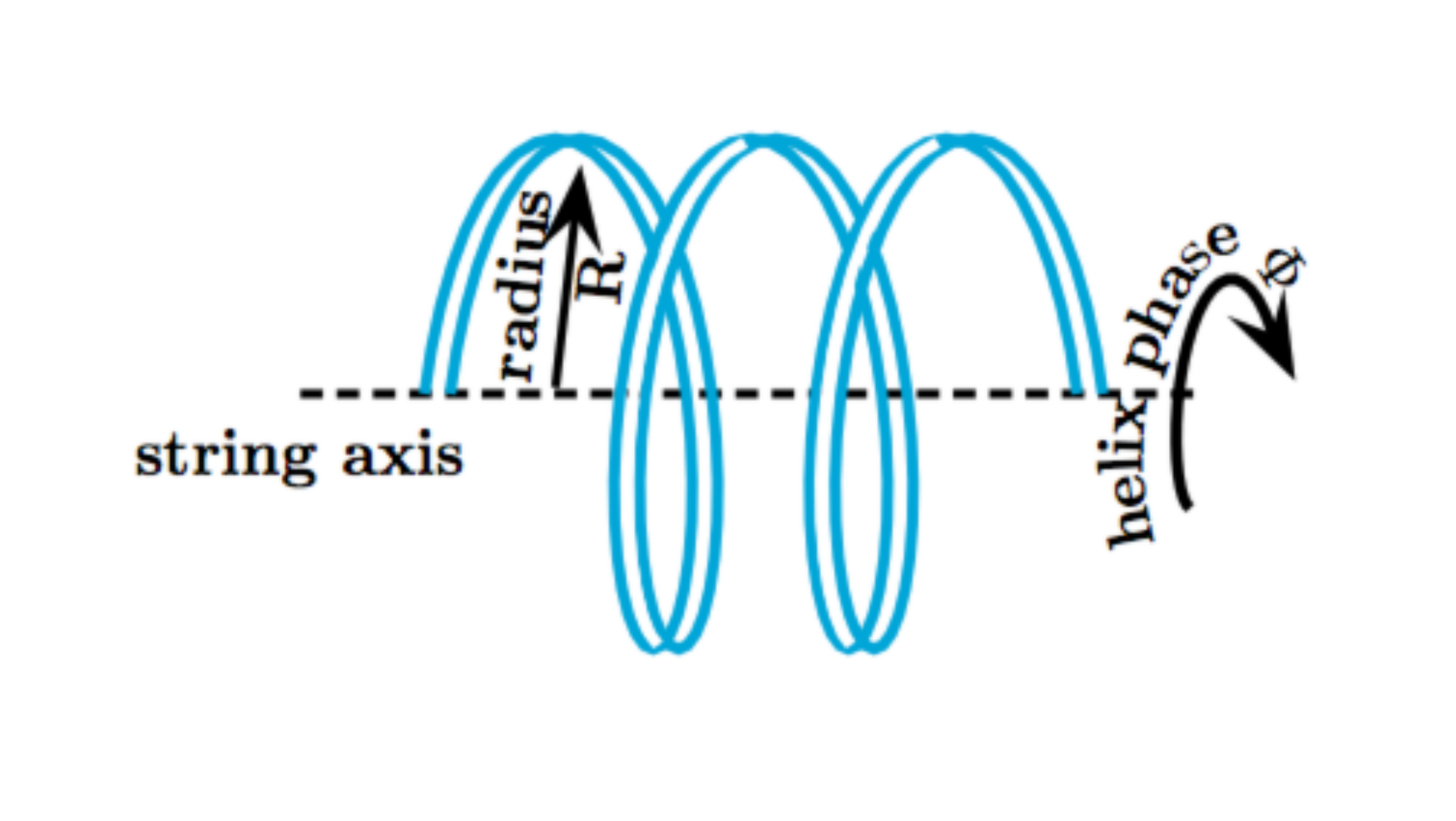} 
  \includegraphics[height=.23\textheight,width=0.45\textwidth]{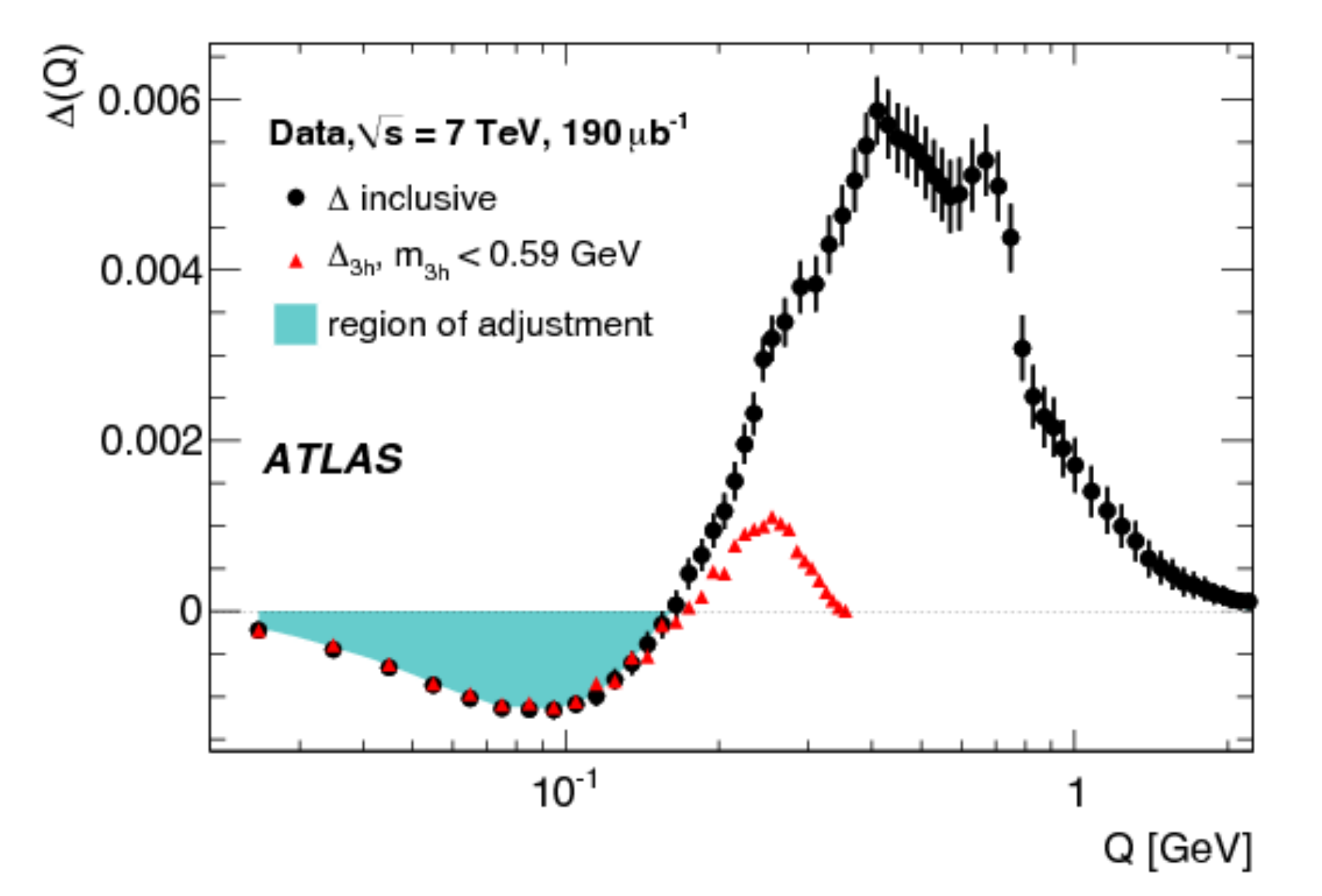}  
  \caption{Left) Parameterization of the helical shape of the QCD string.
    Right) The corrected $\Delta(Q)$ is compared with the corrected contribution
    from low-mass three-hadron chains $\Delta_{\mr 3h}(Q)$. The chain mass limit is
    set to a value of $m_{\mr 3h}^{\mr cut}= 0.59$~GeV, which reproduces the excess
    in the inclusive like-sign pair production at low $Q$ (shaded area). Bin
    errors indicate the combined statistical and reconstruction uncertainty.
    Plots taken from Ref.~\citenum{ATLAS-hadchains2}.}
  \label{chains}
  \end{center}
\end{figure}
Figure~\ref{chains} left represents a parameterization of the helical shape of the
QCD string, while the plot on right showing a ratio of differences between number
of opposite-sign and same-sign charged particles to the number of charged
particles, [N(OS)-N(SS)]/N$_{\mr ch}$, plotted as a function of $Q$, documents that
the low-$Q$ region is described by this new model of fragmentation using helical
strings. The very low-$Q$ region is predicted to be populated exclusively by the
same-sign pairs (which means the rank $r=2$). The further study then showed that
the source of these correlations are 3-hadron chains.

\section{Summary}
This text summarizes recent soft QCD measurements from six LHC experiments and
attempts to document the importance of soft QCD physics in a number of
aspects. For example, the soft QCD results serve as input for modeling pile-up at
LHC and interactions in cosmic rays. They also greatly help in a better
understanding of the hadronization mechanism, typically a mixture of processes
dominating at low transverse momenta. Work on improving hadronization models is
currently ongoing and involves a better treatment of color-confining objects such
as strings and ropes. 
%Very forward flow (also vs central flow) helps to model interactions in cosmic rays
It is also evident that the effect of underlying event is non-negligible at LHC
energies. Finally it was indicated that by studying particle correlations, we
gain not only a powerful tool to learn more about multihadron production, but to
investigate striking similarities between three collision systems at LHC, namely
$PbPb$, $pPb$ and $pp$.
With a limited set of analyses we demonstrated that all collision systems are
useful for soft QCD studies, nicely complementing each other.
Let us also note that the performant LHC machine and experiments unprecedently
provide high-statistics and at the same time high-precision data samples which
enable us to estimate reliably many sources of systematic uncertainties. 
Sophisticated techniques are used to measure very low \pt\ particles, to
efficiently subtract many sources of background and to make use of unfolding
techniques in several dimensions. This way, the precision data help to 
understand unexplained phenomena and to develop or to reject various theoretical
models.
We reported about similar phenomena observed in $PbPb$, $pPb$ and $pp$ (high
multiplicity) collisions, namely about strangeness enhancement and collectivity
effects. The primordial question currently discussed in the HI community is why
these effects are observed in small systems such as $pPb$ and $pp$.
We also conclude that the near-side ridge may be a good testing ground to study
complementarity between hydrodynamics/QGP and dynamics models
(CGC/saturation/ropes). 

\iffalse
\subsection{CME}
Charge-dependent azimuthal correlations serve to measure the so called Chiral magnetic effect (CME) which comes from an interaction of quarks with metastable domains of gluon fields. This interaction leads to imbalance in left- and right-handed quarks which violates local parity symmetry. Non-central nucleus-nucleus collision produce a strong magnetic field and this chirality imbalance leads to an electric current perpendicular to the reaction plane, resulting in a final state charge separation. And that's called CME. 
So what's measured is the 3-particle correlations with respect to 2nd order event plane.  We form same-sign and opposite-sign particle pairs and the 3rd particle in forward calorimeter to probe long-range correlations. These 3-particle correlations are shown as a function of multiplicity and centrality for $pPb$ and $PbPb$ collisions and we observe very similar effects for both collision systems. 
Their difference, namely opposite signs minus same sign again as a function of multiplicity is what is believed to come from the Chiral magnetic effect. What is a surprise is that the same effect is measured both in $PbPb$ and $pPb$, while in $pPb$ it was not expected due to the fact that the magnetic field is smaller than in peripheral $PbPb$ collisions and that the angle between magnetic field and event plane is randomly distributed. But if you look in more detail, you can spot that slopes may not be completely the same. 
\fi

\section{Acknowledgement}
Supported by the project LG15052 of the Ministry of Education, Youth and Sports
of the Czech Republic. Author wishes to thank Edward Sarkisyan-Grinbaum for
helpful discussions.

\iffalse
\section{{\btex}ing}
If you use the \btex\ program to maintain your bibliography, you do
not use the \verb|thebibliography| environment. Instead, you should
include the lines

\begin{verbatim}
\bibliographystyle{ws-procs961x669}
\bibliography{ws-pro-sample}
\end{verbatim}

\noindent where \verb|ws-procs961x669| refers to a file \verb|ws-procs961x669.bst|,
which defines how your references will look.
The argument to \verb|\bibliography| refers to the file
\verb|ws-pro-sample.bib|, which should contain your database in
\btex\ format. Only the entries referred to via \verb|\cite| will be
listed in the bibliography.

\bibliographystyle{ws-procs961x669}
\bibliography{ws-pro-sample}

\begin{thebibliography}{100}
\bibitem{ALICE} ALICE Collab., {\em The ALICE experiment at the CERN LHC},
  {\em JINST} {\bf 3}, S08002 (2008).
\bibitem{ATLAS} ATLAS Collab., {\em The ATLAS Experiment at the CERN Large Hadron Collider}, {\em JINST} {\bf 3}, S08003 (2008). 
\bibitem{CMS} CMS Collab., {\em The CMS Experiment at the CERN LHC}, {\em JINST} {\bf 3}, S08004 (2008).
\bibitem{LHCb} LHCb Collab., {\em The LHCb Detector at the LHC}, {\em JINST} {\bf 3}, S08005 (2008).
\bibitem{LHCf} LHCf Collab., {\em The LHCf detector at the CERN Large Hadron Collider}, {\em JINST} {\bf 3}, S08006 (2008).
\bibitem{Totem} TOTEM Collab., {\em The TOTEM experiment at the CERN Large Hadron Collider}, {\em JINST} {\bf 3}, S08007 (2008).  
\bibitem{compete}  COMPETE Collab., {\em Benchmarks for the forward observables at RHIC, the Tevatron Run II and the LHC}, {\em Phys. Rev. Lett.} {\bf 89}, 201801 (2002).  
\bibitem{epl101-tot} TOTEM Collab., {\em Luminosity-independent measurements of total, elastic and inelastic cross-sections at \sq\ = 7~TeV}, {\em Europhys. Lett.} {\bf 101}, 21004 (2013).
\bibitem{Totem-8TeV-sigmatot} TOTEM Collab., {\em Measurement of Elastic pp
Scattering at $\sqrt{s}$ = 8~TeV in the Coulomb-Nuclear Interference Region
- Determination of the $\rho$ Parameter and the Total Cross-Section},
  {\em Eur. Phys. J.} {\bf C 76}, 661 (2016).
\bibitem{Totem-2.76TeV-sigmatot} M. Deile for the TOTEM Collab., talk at the
  conference EDS Blois 2017, Prague, Czech Rep.
\bibitem{ATLAS-8TeV-sigmatot} ATLAS Collab., {\em Measurement of the total cross section from elastic scattering in pp collisions at \sq\ = 8~TeV with the ATLAS detector}, {\em Phys. Lett.} {\bf B 761}, 158  (2016).
\bibitem{ATLAS-13TeV-inel} ATLAS Collab., {\em Measurement of the Inelastic Proton-Proton Cross Section at \sq\ = 13~TeV with the ATLAS Detector at the LHC},
  {\em Phys. Rev. Lett.} {\bf 117}, no.18, 182002 (2016).
\bibitem{PYTHIA8} T. Sj\"{o}strand, S. Mrenna, and P. Z. Skands, {\em A Brief Introduction to PYTHIA 8.1}, {\em Comput. Phys. Commun.} {\bf 178}, 852 (2008).
\bibitem{EPOS} T. Pierog {\em et al.}, {\em EPOS LHC: Test of collective hadronization
  with data measured at the CERN Large Hadron Collider}, {\em Phys. Rev.} {\bf C 92} no. 3, 034906 (2015).
\bibitem{QGSJETII} S. Ostapchenko, {\em Monte Carlo treatment of hadronic interactions in enhanced Pomeron scheme: I. QGSJET-II model}, {\em Phys. Rev.}
  {\bf D 83}, 014018 (2011).
\bibitem{Totem-nonexp} TOTEM Collab., {\em Evidence for Non-Exponential Elastic
  Proton-Proton Differential Cross-Section at Low $|t|$ and \sq\ = 8~TeV by
  TOTEM}, {\em  Nucl. Phys.} {\bf B 899}, 527 (2015).  
\bibitem{ALICE-incl-ch} ALICE Collab., {\em Charged-particle multiplicities in
  proton-proton collisions at $\sqrt{s}$ = 0.9 to 8~TeV}, {\em Eur. Phys. J.}
  {\bf C 77}, 33 (2017). 
\bibitem{ATLAS-13TeV-incl-ch} ATLAS Collab., {\em Charged-particle distributions at low transverse momentum in \sq\ = 13~TeV pp interactions measured with the
  ATLAS detector at the LHC} {\em Eur. Phys. J.} {\bf C 76}, 502 (2016).
\bibitem{CMS-13TeV-CASTOR} CMS Collab., {\em Measurement of the inclusive energy
  spectrum in the very forward direction in proton-proton collisions at \sq = 13~TeV}, {\em J. High En. Phys.} {\bf 08}, 046 (2017).
\bibitem{LHCf-7TeV-n} LHCf Collab., {\em Measurement of very forward neutron energy spectra for 7~TeV proton–proton collisions at the Large Hadron Collider},
  {\em Phys. Lett.} {\bf B 750}, 360 (2015).
\bibitem{DPMJET} F. W. Bopp {\em et al.}, {\em Antiparticle to Particle Production Ratios in Hadron-Hadron and d-Au Collisions in the DPMJET-III Monte Carlo}, {\em Phys. Rev.} {\bf C 77},
  014904 (2008).
\bibitem{SYBILL}  E.-J. Ahn {\em et al.}, {\em Cosmic ray interaction event generator SIBYLL 2.1}, {\em Phys. Rev.} {\bf D 80}, 094003
  (2009).
\bibitem{LHCf-7TeV-pi0} LHCf Collab., {\em Measurements of longitudinal and transverse momentum distributions for neutral pions in the forward-rapidity region with the LHCf detector}, {\em Phys. Rev.} {\bf D 94}, 032007 (2016).
\bibitem{LHCf-13TeV-gamma} LHCf Collab., {\em Measurement of forward photon-energy spectra for \sq\ = 13~TeV proton-proton collisions with the LHCf detector},
  {\em CERN-EP-2017-051, arXiv:1703.07678 [hep-ex]}.
\bibitem{ALICE-id} ALICE Collab., {\em Enhanced production of multi-strange hadrons in high-multiplicity proton-proton collisions}, {\em Nat. Phys.} {\bf 13},
  535 (2017).
  \bibitem{strangeHI} P. Koch, B. Muller and J. Rafelski, {\em Strangeness in Relativistic Heavy Ion Collisions}, {\em Phys. Rept.} {\bf 142}, 167 (1986).
\bibitem{DIPSY}  C. Bierlich and J. R. Christiansen, {\em Effects of Colour Reconnection on Hadron Flavour Observables}, {\em Phys. Rev.} {\bf D 92}, 094010 (2015).
  
\bibitem{CMS-13TeV-id} CMS Collab., {\em Measurement of charged pion, kaon, and proton production in proton-proton collisions at \sq\ = 13~TeV},
  {\em CERN-EP-2017-091, arXiv:1706.10194 [hep-ex]}.  
\bibitem{ATLAS-13TeV-UE} ATLAS Collab., {\em Measurement of charged-particle distributions sensitive to the underlying event in \sq\ =13~TeV proton--proton collisions with the ATLAS detector at the LHC }, {\em J. HEP} {\bf 03}, 157 (2017).
\bibitem{ATLAS-097TeV-UE} ATLAS collab., {\em Measurement of underlying event characteristics using charged particles in pp collisions at \sq\ = 900~GeV and 7~TeV with the ATLAS detector}, {\em Phys.Rev.} {\bf D 83}, 112001 (2011). 
\bibitem{CMS-13TeV-UE} CMS Collab., {\em Measurement of the underlying event using the Drell-Yan process in proton-proton collisions at \sq\ = 13~TeV},
  {\em CERN-EP-2017-249, arXiv:1711.04299 [hep-ex]}. 
\bibitem{M+R} M. Sumbera and R. Pasechnik, {\em Phenomenological Review on Quark–Gluon Plasma: Concepts vs. Observations}, {\em Universe} {\bf3}, no.1, 7 (2017).
\bibitem{CGC} L.D. McLerran and R. Venugopalan, {\em Computing quark and gluon distribution functions for very large nuclei}, {\em Phys. Rev.} {\bf D 49},
  2233 (1994).
\bibitem{Bjorken_ridge} J. D. Bjorken {\em et al.}, {\em Possible multiparticle ridge-like correlations in very high multiplicity proton-proton collisions}, {\em Phys. Lett.} {\bf B 726}, 344 (2013).  
\bibitem{LHCb-5TeV-dphi} LHCb Collab., {\em Measurements of long-range near-side
  angular correlations in \sq\ = 5.02~TeV proton-lead collisions in the forward region}, {\em Phys. Lett.} {\bf B 762}, 473 (2016).
\bibitem{ALICE-5TeV-dphi} ALICE Collab., {\em Evolution of the longitudinal and azimuthal structure of the near-side jet peak in Pb-Pb collisions at \sq\ =2.76~TeV}, {\em Phys. Rev.} {\bf C 96}, no.3,
  034904 (2017).
\bibitem{ATLAS-3subev} ATLAS Collab., {\em Measurement of multi-particle azimuthal correlations with the subevent cumulant method in pp and p+Pb collisions with the ATLAS detector at the LHC}, {\em CERN-EP-2017-160, arXiv:1708.03559 [hep-ex]}.
\bibitem{CMS-corrnch} CMS Collab., {\em Measurement of Long-Range Near-Side Two-Particle Angular Correlations in pp Collisions at \sq\ = 13~TeV},
  {\em Phys. Rev. Lett.} {\bf 116}, 172302 (2016).
\bibitem{ATLAS-v2nch} ATLAS Collab., {\em Measurement of multi-particle azimuthal correlations in pp, p+Pb and low-multiplicity Pb+Pb collisions with the ATLAS
  detector}, {\em Eur. Phys. J.} {\bf C 77}, 428 (2017).
\bibitem{CMS-multipc} CMS Collab., {\em Evidence for collectivity in pp collisions at the LHC }, {\em Phys. Lett.} {\bf B 765}, 193 (2017).
\bibitem{CMS-mpc} CMS Collab., {\em Multiplicity and transverse momentum dependence of two- and four-particle correlations in pPb and PbPb collisions},
  {\em Phys. Lett.} {\bf B 724}, 213 (2013).    
\bibitem{ALICE-corrid} ALICE Collab., {\em Insight into particle production mechanisms via angular correlations of identified particles in pp collisions at \sq\ = 7~TeV}, {\em Eur. Phys.J.} {\bf C 77}, no.8, 569 (2017).  
\bibitem{ATLAS-7TeV-pp-BEC} ATLAS Collab., {\em Two-particle Bose--Einstein correlations in pp collisions at \sq\ = 0.9 and 7~TeV measured with the ATLAS detector}, {\em Eur. Phys. J} {\bf C 75}, 466 (2015).
\bibitem{CMS-BEC} CMS Collab., {\em Measurement of Bose-Einstein Correlations in pp Collisions at \sq\ =0.9 and 7~TeV}, {\em J. High Energy Phys.} {\bf 05}, 029
  (2011).
\bibitem{UA1-BEC} UA1 Collab., {\em Bose-Einstein Correlations in $\bar{p}p$ Interactions at \sq\ = 0.2 to 0.9~TeV}, {\em Phys. Lett.} {\bf B 226}, 410 (1989).
\bibitem{ATLAS-pPb-BEC} ATLAS Collab., {\em Femtoscopy with identified charged pions in proton-lead collisions at \sq\ =5.02~TeV with ATLAS}, {\em CERN-EP-2017-004, arXiv:1704.01621 [hep-ex]}. 
\bibitem{LHCb-BEC} LHCb Collab., {\em Bose-Einstein correlations of same-sign charged pions in the forward region in pp collisions at \sq\ = 7~TeV},
  %{\em CERN-EP-2017-203, arXiv:1709.01769 [hep-ex]}.
  {\em J. High Enery Phys. {\bf 1712} 025 (2017)}.
\bibitem{ALICE-pion-BEC} ALICE Collab., {\em Multipion Bose-Einstein correlations in pp,p-Pb, and Pb-Pb collisions at energies available at the CERN Large Hadron Collider}, {\em Phys. Rev.} {\bf C 93}, 054908 (2016).
\bibitem{Sarka} S. Todorova-Nova, {\em Quantization of the QCD string with a helical structure}, {\em Phys. Rev.} {\bf D 89}, no.1, 015002 (2014).
\bibitem{helix} B. Andersson {\em et al.}, {\em Is there screwiness at the end of the
  QCD cascades?}, {\em J. High Energy Phys.} {\bf 9809}, 014 (1998).
\bibitem{ATLAS-hadchains2} ATLAS Collab., {\em Study of ordered hadron chains with the ATLAS detector}, {\em Phys. Rev. {\bf D 96}, no.9, 092008 (2017)}.  

\end{thebibliography}

\fi

%\end{document}

%Non BiBTeX users can list down their references as:

\end{document}